# Bayesian Multivariate Track Geometry Degradation Modelling and its use in Condition-Based Inspection

Huy Truong-Ba[1,*], Sinda Rebello[2], Michael E. Cholette[1], Venkat Reddy[2], Pietro Borghesani[3]

*1. Science and Engineering Faculty, Queensland University of Technology, Brisbane Queensland, Australia*

*2. Queensland Rail, Brisbane, Australia*

*3. School of Mechanical and Manufacturing Engineering, UNSW Sydney, Australia*

**Corresponding author:* h.truongba@qut.edu.au

## Abstract

Effective maintenance of railway infrastructure is crucial for safe and comfortable transportation. Among the various degradation modes, track geometry deformation due to repeated loading significantly impacts operational safety. Detecting and maintaining acceptable track geometry involves the use of track recording vehicles (TRVs) that inspect and record geometric parameters. This study aims to develop a novel track geometry degradation model that considers multiple indicators and their correlations, accounting for both imperfect manual and mechanized tamping. A multivariate Wiener model is formulated to capture the characteristics of track geometry degradation. To address data limitations, a hierarchical Bayesian approach with Markov Chain Monte Carlo (MCMC) simulation is employed. This research contributes to the analysis of a multivariate predictive model, which considers the correlation between the degradation rates of multiple indicators, providing insights for rail operators and new track-monitoring systems. The model's performance is validated through a real-world case study on a commuter track in Queensland, Australia, using actual data and independent test datasets. Additionally, the study demonstrates the application of the proposed multivariate degradation model in developing a condition-based inspection policy for track geometry, potentially reducing the number of TRV runs while maintaining abnormal detection levels and failure rates.

## 1 Introduction

Railways span thousands of kilometers, representing a costly asset with significant operational and maintenance (O&M) challenges. Like any infrastructure, rail tracks degrade over time, requiring effective maintenance to ensure safe and comfortable transport. A key degradation mechanism is track geometry deformation due to repeated loading, which directly impacts train safety and demands substantial maintenance resources [1–5].

Track geometry degradation is typically monitored through inspections using track recording vehicles (TRVs), which measure parameters such as longitudinal level, alignment, gauge, cant, and twist [6,7]. When anomalies are detected, the affected track segment is scheduled for maintenance using specialized tamping vehicles [3,8]. Given the vast network and limited resources (TRVs, tamping vehicles, and crews), optimizing inspection and maintenance scheduling is a significant challenge. Effective planning must balance the risk of track defects against the cost of frequent inspections, underscoring the need for predictive models to track and forecast degradation.

This challenge has driven research into degradation and reliability modeling, particularly considering rail networks' spatial distribution and uncertainty factors (e.g., environmental effects, tamping) [6]. In general, track geometry degradation models can be categorized into two main types: mechanical models and (data-driven) statistical models [6,9,10]. Mechanical models rely on physical relationships among track components but are limited by their deterministic nature, requiring detailed system knowledge (e.g., soil conditions, loading) [9]. Indeed, statistical approaches have become increasingly popular, owing to their ability to both bypass the need for a detailed knowledge of network parameters/loadings, and to provide insight into the impact of uncertain factors (e.g. environmental factors, operating conditions) [11–16]. These data-driven approaches depend on historical track geometry data to fit statistical degradation trends [13,17] and/or quantify the uncertain effects on predicted degradation indicator [16,18], which are then expected to repeat in the future. Recently, some studies have introduced prediction models for track geometry based on modern artificial intelligence (AI) techniques such as neuron network (NN) [19,20]. However, these NN methods have limitations, including a the need for a large amount of track geometry data, difficulty in incorporating uncertainty, and difficulty in interpreting the resulting model [19].

Statistical studies often use Gamma or Wiener processes to model historical TRV data and predict future degradation [21,22]. While Gamma models have been applied to track degradation [12,23], their monotonic nature makes them unsuitable for track geometry parameters, which exhibit random, non-monotonic (self-healing) behavior. Consequently, the Wiener process is more widely used [11,24–26]. While physically a continuous process, track geometry parameters are only sampled during TRV inspections, and thus often modelled using discrete condition state transitions [27–29].

Most existing models consider individual track indicators, often focusing on longitudinal level as the primary measure [29]. However, maintenance decisions are based on the worst-performing indicator, necessitating models that account for multiple parameters. Moreover, given the physical relationships between different indicators (and often even their mathematical expression) it is important for an accurate stochastic model to consider the possibility of inter-relationships among the different quantities being monitored. Few studies have explored multivariate degradation. Mercier et al. [23] developed a track geometry degradation model with longitudinal and transversal (alignment) leveling indicators according to a bivariate Gamma process, while Arasteh khouy I et al. [30] incorporated longitudinal, lateral deviations, and cant errors. Some studies have used track quality indices to reduce multivariate complexity [15,31,32], but this approach, like single-indicator models, fails to capture correlations among track indicators.

Another potential problem of modelling track geometry degradation is data sparsity. With limited TRV runs per track segment, parameter estimation methods (e.g., maximum likelihood) may be less effective. Bayesian approaches are a typical paradigm for understanding the uncertainty of parameter estimation when data is sparse [33]; yet few studies have explored them. Andrade and Teixeira [34] developed a linear degradation model for track longitudinal indicator in which random coefficients were estimated via Bayesian approach with a Markov Chain Monte Carlo (MCMC) simulation. Extended works from the same authors [35,36] modified the linear degradation model considering tamping and renewal decisions. The Hierarchical Bayesian approach was used to estimate parameters of the modified model. However, these works addressed only the univariate degradation process of the track longitudinal parameter.

This study develops a track geometry degradation model that accounts for multiple indicators, their correlations, and the impact of imperfect tamping. A multivariate Wiener process is formulated to capture these characteristics, and a hierarchical Bayesian approach with MCMC simulation is applied to address data limitations. This study therefore makes three significant contributions to existing literature:

- **Novel correlation analysis and multivariate predictive model:** This is the first analysis to investigate the correlation among track geometry indicators and incorporate these findings into a multivariate predictive model. The correlation analysis itself is a novel and significant contribution, offering valuable insights for rail operators and researchers. It could inform the development of new track-monitoring systems, such as requirements for onboard track monitoring. By comparing the developed multivariate degradation model with univariate models, this study addresses the research question of whether inspection and maintenance policies based on single-indicator degradation models provide accurate and beneficial decisions.
- **Methods to address data limitations:** The study introduces two rigorous solutions to address common data limitations: a novel statistical pre-processing method for identifying tamping events and a parameter estimation methodology based on the hierarchical Bayesian paradigm.

- **Validation with real data and deployment in a Condition-Based Inspection Policy:** The model is validated using real-world data from a commuter track in Queensland, Australia. The models were estimated using actual data, with predictions benchmarked against a test dataset not used in model calibration. Furthermore, a condition-based inspection policy is developed based on the multivariate degradation model, demonstrating practical application even in the absence of comprehensive cost information.

The remainder of this paper is organized as follows: Section 2 gives a detailed theoretical background of the application of a multivariate Wiener model for track degradation and the corresponding parameter estimation using a hierarchical Bayesian technique. Section 3 describes the case study and data pre-processing to identify tamping events via statistical analysis. The results obtained with the new multi-variate degradation model and their benchmark against single indicator models are presented in Section 4. Section 5 discuses an application of developed multivariate degradation model on developing a condition-based inspection policy for the real case study under the practice of lacking cost information. The last section provides a summary of the contributions as well as potential extensions of this study.

# 2 Geometry degradation model and Bayesian fitting

This study develops a track geometry degradation model incorporating multiple indicators (e.g., longitudinal and alignment parameters) and benchmarks it against a single indicator model. Track geometry distortion, measured by deviations from designed values, is not monotonic over time as illustrated by real data in Section 3.2. Thus, typical monotonic models like the Gamma process are unsuitable, whereas the Wiener process [21] is more appropriate. Additionally, the Wiener process is convenient for modelling and analyzing multiple variables (geometry indicators). This work employs a multivariate Wiener process [37] to model track geometry degradation and multiple geometry indicators. In the next subsection such multi-variate Wiener model is introduced and compared to the traditional univariate model used in literature to describe the evolution of single indicators independently. In the second subsection of this chapter, the hierarchical Bayesian approach used for the fitting of the model parameters is described.

## 2.1 Wiener models

Track geometry indicators are collected by TRV at each inspection, with a resolution of 250 mm per record. Yet, direct analysis at this resolution is challenging because of the difficulty in aligning records from different TRV runs and the large amount of data. A common approach is to segment the track usually 100 or 200 meters in length [6,16,19] and this approach is adopted in this study. Records from TRV for each segment are then summarized using statistical quantities (e.g. maximum absolute deviation). These values are subsequently used for degradation analyses and modelling.

Let $Z_{q,i}(t) \in \mathbb{R}$ be a random geometry indicator value of indicator $q$ of track segment $i$ at time $t$. All the available information on the geometry of segment $i$ at time $t$ is thus summarized by the vector $\mathbf{Z}_i(t) \in \mathbb{R}^{N_q}$ of all $N_q$ geometry indicators, i.e.:

$$\mathbf{Z}_i(t) = \left[Z_{1,i}(t), Z_{2,i}(t), \dots, Z_{N_q,i}(t)\right]^T \tag{1}$$

A multivariate Wiener process is used to model the evolution of $\mathbf{Z}_i(t) = \boldsymbol{\mu}_i t + \mathbf{L}_i \mathbf{B}(t)$, where $\boldsymbol{\mu}_i \in \mathbb{R}^{N_q}$ is the vector containing the process drifts , $\mathbf{L}_i$ is the Cholesky factor of a positive semi-definite covariance matrix $\boldsymbol{\Sigma}_i = \mathbf{L}_i \mathbf{L}_i^T \in \mathbb{R}^{N_q \times N_q}$ and $\mathbf{B}(t) \in \mathbb{R}^{N_q}$ is a vector of $N_q$ independent standard Brownian motions. In practice, a finite and discrete dataset $\widehat{\mathcal{D}}_i = \{\mathbf{z}_{i,k} | k = 1,2, \dots, K\}$ of measurements of $\mathbf{Z}_i(t)$ are available at specific times $t_{i,k}$, $k = 1, \dots, K_i$, with $\mathbf{z}_{i,k} = \mathbf{Z}_i(t_{i,k})$, and therefore $z_{q,i,k} = Z_{q,i}(t_{i,k})$. Under these modelling assumptions, the variation of indicators between two successive measuring times $t_{i,k-1}$ and $t_{i,k}$ follows a multivariate Normal distribution:

$$\Delta \mathbf{z}_{i,k} \sim \mathcal{N}\left(\boldsymbol{\mu}_i \Delta t_{i,k}, \boldsymbol{\Sigma}_i \Delta t_{i,k}\right) \tag{2}$$

where $\Delta t_{i,k} = t_{i,k} - t_{i,k-1}$ and $\Delta \mathbf{z}_{i,k} = \mathbf{z}_{i,k} - \mathbf{z}_{i,k-1}$.

The multi-variate model discussed so far is an extension of the literature univariate case. The latter can be obtained by constraining the covariance matrices $\boldsymbol{\Sigma}_i$ to be diagonal. Under this special case, the degradation of the indicators is modelled as a series of independent univariate Wiener processes. In this case, (2) simplifies to:

$$z_{q,i,k} - z_{q,i,k-1} \sim \mathcal{N}\left(\mu_{q,i,k} \cdot \Delta t_{i,k}, \sigma_{q,i}^2 \cdot \Delta t_{i,k}\right) \tag{3}$$

with $\sigma_{q,i}^2$ representing the diagonal (and only non-zero) terms of the univariate covariance matrix, i.e., the variances of the univariate Wiener processes.

The track geometry is restored when tamping (i.e. maintenance) activities are conducted (**Fig. 1**). Tamping is modelled as an imperfect maintenance action which restores the track geometry indicators to some values between the values before tamping and the design targets. However, it is common that the tamping actions are not recorded with sufficient location specificity and time resolution to identify the exact tamping time for a specific segment. Therefore, if tamping occurs during the $k$-th inspection interval $\left(t_{i,k-1}, t_{i,k}\right]$ for segment $i$ it is assumed to occur at the midpoint of the time interval.

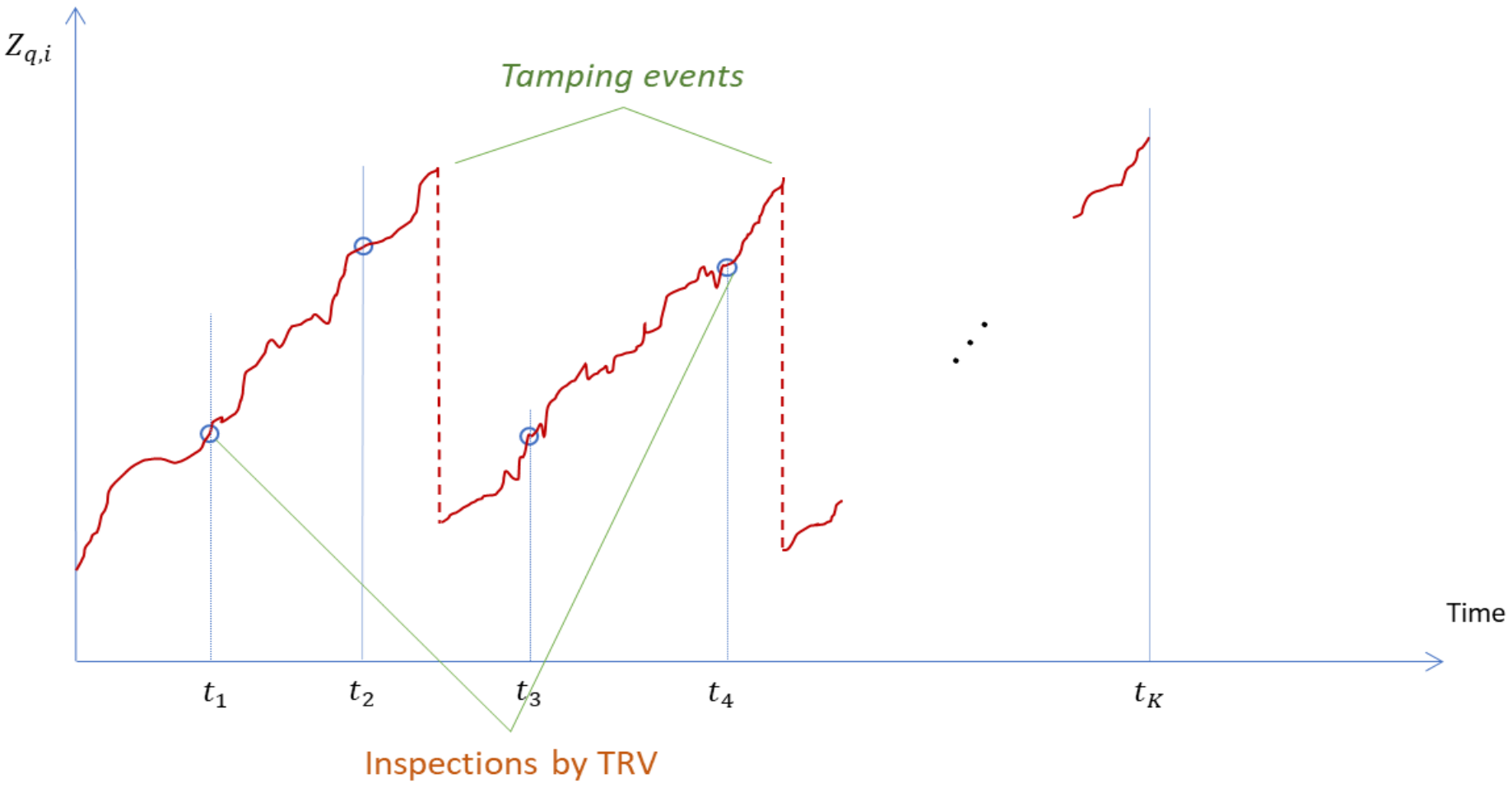

**Fig. 1** Illustration of track geometry degradation and tamping

Since the maintenance events happen only in some inspection intervals, the time intervals are classified into two complementary sets: one composed of interval indices with maintenance ($\mathbb{M}_i$) and a second composed of those without maintenance $\overline{\mathbb{M}}_i$. Let $\delta_{i,k}$ be the indicator function of maintenance events, $\delta_{i,k} = 1$ if the maintenance happens in interval $\left(t_{i,k-1}, t_{i,k}\right]$, and $\delta_{i,k} = 0$ otherwise. The sets $\mathbb{M}_i$ and $\overline{\mathbb{M}}_i$ are defined as follows:

$$\mathbb{M}_i = \left\{k \middle| \delta_{i,k} = 1\right\} \quad \text{and} \quad \overline{\mathbb{M}}_i = \left\{k \middle| \delta_{i,k} = 0\right\} \tag{4}$$

Let $\mathbf{z}_{i,k}^{+} = \left\{z_{q,i,k}^{+} \middle| q = 1,2, \dots, N_q, k \in \mathbb{M}_i\right\}$ and $\mathcal{D}_i$be the vector of geometry indicators just after the tamping event in the interval $k \in \mathbb{M}_i$ (assumed to be at $t_{i,k-1} + \Delta t_{i,k}/2$ for an interval including maintenance). For this interval, maintenance will therefore alter the normal degradation process of Eq. (2), resulting in the track geometry indicators at $t_{i,k}$ (the end of interval):

$$\mathbf{Z}_i\left(t_{i,k}\right) \sim \mathcal{N}\left(\mathbf{z}_{i,k}^{+} + \boldsymbol{\mu}_i \frac{\Delta t_{i,k}}{2}, \boldsymbol{\Sigma}_i \frac{\Delta t_{i,k}}{2}\right) \quad \forall k \in \mathbb{M}_i \tag{5}$$

The condition at the end of the interval is no longer dependent on the condition at the beginning of the interval, but rather to the additional parameters $\mathbf{z}_{i,k}^{+}$.

This additional parameter is therefore present in the likelihood functions of both models, which show the combined contributions of intervals with maintenance ($\mathbb{M}_i$) and without ($\overline{\mathbb{M}}_i$). The multivariate Wiener model has likelihood:

$$
\begin{aligned}
&\mathcal{L}\left(\boldsymbol{\mu}_i, \boldsymbol{\Sigma}_i, \left\{\mathbf{z}_{i,k}^{+} \middle| k \in \mathbb{M}_i\right\}; \mathcal{D}_i\right) \\
&\qquad = \prod_{k \in \overline{\mathbb{M}}_i} \frac{\exp\left[-\frac{1}{2}\left(\Delta\mathbf{z}_{i,k} - \boldsymbol{\mu}_i \Delta t_{i,k}\right)^T \left(\Delta t_{i,k} \boldsymbol{\Sigma}_i\right)^{-1} \left(\Delta\mathbf{z}_{i,k} - \boldsymbol{\mu}_i \Delta t_{i,k}\right)\right]}{\sqrt{\left(2\pi \Delta t_{i,k}\right)^{N_q} |\boldsymbol{\Sigma}_i|}} \\
&\qquad \cdot \prod_{k \in \mathbb{M}_i} \frac{\exp\left[-\frac{1}{2}\left(\mathbf{z}_{i,k} - \mathbf{z}_{i,k}^{+} - \frac{\boldsymbol{\mu}_i \Delta t_{i,k}}{2}\right)^T \left(\frac{\Delta t_{i,k} \boldsymbol{\Sigma}_i}{2}\right)^{-1} \left(\mathbf{z}_{i,k} - \mathbf{z}_{i,k}^{+} - \frac{\boldsymbol{\mu}_i \Delta t_{i,k}}{2}\right)\right]}{\sqrt{\left(\pi \Delta t_{i,k}\right)^{N_q} |\boldsymbol{\Sigma}_i|}}
\end{aligned}
\tag{6}
$$

where the first product is the standard likelihood of multivariate Normal distribution for case without tamping and second product is the likelihood of case with tamping.

## 2.2 Bayesian Fitting

With the parameter $\mathbf{z}_{i,k}^{+}$ in the track geometry degradation model described above, the fitting parameters via traditional Maximum Likelihood Estimation (MLE) approach is difficult. MLE is most naturally applied to fitting data from each segment independently, but maintenance events are sparse — many segments are maintained only once or not at all. It is therefore advantageous to use pre- and post-maintenance data from other segments to estimate $\mathbf{z}_{i,k}^{+}$ via some sort of pre-processing step prior to estimating other parameters (e.g. the mean value of $\mathbf{z}_{i,k}^{+}$ over all or some subset of segments). Yet, selecting this pre-estimation comes with the danger of significantly biasing the subsequent parameter estimation. A second challenge with MLE method is the need for sufficient data for getting reliable parameter estimates. This is particularly problematic for $\mathbf{z}_{i,k}^{+}$, where data will only be available for some segments. These obstacles can be overcome by using Bayesian fitting techniques [38], in which, the parameter $\mathbf{z}_{i,k}^{+}$ will be estimated together with other parameters $\boldsymbol{\mu}_i$, $\boldsymbol{\Sigma}_i$ of Wiener model.

For a specific segment $i$, let $\theta_i = \{\mathbf{z}_{i,k}^{+}, \boldsymbol{\mu}_i, \boldsymbol{\Sigma}_i\}$ be the parameter set need to be estimated via dataset $\mathcal{D}_i = \left\{\mathbf{z}_{i,k} \middle| k = 1,2,\dots,K\right\}$. The prior of $\theta_i$is denoted as $p(\theta_i)$ and the posterior $p(\theta_i|\mathcal{D}_i)$ has relation to the prior according to Bayesian rule:

$$
p(\theta_i|\mathcal{D}_i) = \frac{\ell(\mathcal{D}_i|\theta_i)\, p(\theta_i)}{\int \ell(\mathcal{D}_i|\theta_i)\, p(\theta_i) d\theta_i} \quad \text{or} \quad p(\theta_i|\mathcal{D}_i) \propto \ell(\mathcal{D}_i|\theta_i)\, p(\theta_i)
$$

The algebraic calculation of term $\int \ell(\mathcal{D}_i|\theta_i)\, p(\theta_i) d\theta_i$ is almost impossible, and therefore, the numerical integration can be considered, but this approach has a number of computational disadvantages. Another powerful way to obtain the posterior with any prior distribution is using Markov Chain Monte Carlo (MCMC) methods [38], which is implemented in standard software such as PyMC (Python) [39], Stan or R [38]. The log-likelihood $\ell(\mathcal{D}_i|\theta_i)$ is the logarithm of equation (6) or

**Error! Reference source not found.** according to multivariate or univariate Wiener models, respectively.

For the multivariate Wiener approach proposed in this study, 2+$|\mathbb{M}_i|$ vectors of parameters, namely $\boldsymbol{\mu}_i, \boldsymbol{\Sigma}_i$ and $\{\mathbf{z}_{i,k}^{+} | k \in \mathbb{M}_i\}$, need to be estimated for each segment model, with $|\mathbb{M}_i|$ representing the cardinality of $\mathbb{M}_i$, i.e., the number of maintenance actions performed on segment $i$.

The drift parameters of priors are set to be half-normal priors, which are assumed to be independent across different geometry indicators:

$$\mu_{i,q} \sim \mathcal{H}alf\mathcal{N}orm(s_q^{\mu}) \tag{7}$$

Such distribution has positive support, ensuring a drift with an expected value corresponding to actual degradation, while the possibility of random self-healing events is left to the subsequent Wiener processes. The covariance matrices $\boldsymbol{\Sigma}_i$ of the Wiener processes must be positive semidefinite and symmetric. Therefore, the prior of $\boldsymbol{\Sigma}_i$ is constructed according to the Lewandowski-Kurowicka-Joe (LKJ) distribution presented in article [40]. Following this approach, $\boldsymbol{\Sigma}_i$ is obtained by combining two matrices. The first is a diagonal matrix with terms $\sigma_{q,i}$, i.e., the marginal standard deviations governing the scale of the variation of each geometry indicator. The correlation matrix $\mathbf{R}_i \in \mathbb{R}^{N_q \times N_q}$ (symmetric, positive definite with ones on the diagonal) is instead responsible for rendering the process multi-variate (correlation structure). Mathematically, the covariance matrix is therefore obtained as:

$$\boldsymbol{\Sigma}_i = \begin{bmatrix} \sigma_{1,i} & 0 & \cdots & 0 \\ 0 & \sigma_{2,i} & \cdots & 0 \\ \vdots & \vdots & \ddots & \vdots \\ 0 & 0 & \cdots & \sigma_{N_q,i} \end{bmatrix} \mathbf{R}_i \begin{bmatrix} \sigma_{1,i} & 0 & \cdots & 0 \\ 0 & \sigma_{2,i} & \cdots & 0 \\ \vdots & \vdots & \ddots & \vdots \\ 0 & 0 & \cdots & \sigma_{N_q,i} \end{bmatrix} \tag{8}$$

The marginal standard deviations $\sigma_{q,i}$ are sampled using Half Normal priors, ensuring positive-only outputs:

$$\sigma_{q,i} \sim \mathcal{H}alf\mathcal{N}orm(s_q^{\sigma}) \tag{9}$$

A suitable prior for $\mathbf{R}_i$ this is the LKJ distribution has the PDF:

$$p(\mathbf{R}_i) = c_d |\mathbf{R}_i|^{\eta-1} \tag{10}$$

Where $c_d$ is a normalizing constant and $\eta$ is a free parameter with the following cases:

- $\eta = 1$ will result in $\mathbf{R}_i$ being sampled uniformly from the space of all covariance matrices.
- $\eta > 1$ has correlations centered around zero.

- $0 < \eta < 1$ favors correlation, with both positive and negative correlations favored equally.

For the post-maintenance term $\mathbf{z}_{i,k}^{+}$, a Log-Normal distribution was chosen, common to all maintenance events on all segments, and with independent distributions across the different geometry indicators:

$$z_{q,i,k}^{+} \sim \mathcal{LN}\left(m_q^z, s_q^z\right) \tag{11}$$

Like the half-normal, the choice of the log-normal ensures that $z_{q,i,k}^{+}$ is greater than zero. However, differently from the half-normal, the log-normal allows to set the position of the mean independently on the variance, and to position the mode above zero, a necessary characteristic for incorporating prior knowledge regarding maintenance.

For the Bayesian fitting, the parameters $s_q^{\mu}$ in Eq. (7), $s_q^{\sigma}$ in Eq. (9), $\eta$ in Eq. (10) and $m_q^z, s_q^z$ in Eq. (11) are set as constant values according to knowledge from maintenance experts of rail operators.

## 2.3 Hierarchical Bayesian Fitting

Hierarchical Bayesian estimation is a powerful method for fitting data-driven stochastic degradation models, particularly with limited data. One of the significant advantages of the hierarchical Bayesian approach is that it is well suited for spatial data, such as data from linear assets like rail tracks, roads, or electricity lines. The general framework of Hierarchical Bayesian model is described as follows. Let $y_i$ be observation (i.e. data) generated from its stochastic model with parameter $\theta_i$, $p(y_i|\theta_i)$. All parameters $\theta_i$ are assumed following a common prior with parameter $\phi$, $p(\theta_i|\phi)$. The parameter $\phi$ of prior is called as hyperparameter and it is also generated from a hyperprior distribution $p(\phi)$. According to Bayes' law, the posterior can be written as:

$$p(\theta_i|y_i) \propto p(y_i|\theta_i)p(\theta_i) = p(y_i|\theta_i)p(\theta_i|\phi)p(\phi)$$

In this hierarchical Bayesian fitting, data of systems with similar degradation (i.e., track segments) is "pooled" to enrich the data for fitting, thereby decreasing the estimated parameter variance. This pooling property is controlled by hyperprior distributions, which represent "global" distributions of model parameters for all segments and all maintenance events. The Markov Chain Monte-Carlo (MCMC) simulation with sampling processes from hyperprior $p(\phi)$ and prior $p(\theta_i|\phi)$ can also be applied to estimate the posteriors.

In the hierarchical Bayesian approach for the track geometry degradation, the statistical models governing specific segments and maintenance events (through $\theta_i = \{\mathbf{z}_{i,k}^{+}, \boldsymbol{\mu}_i, \boldsymbol{\Sigma}_i\}$) are seen as particular "realizations" of the hyperprior distributions, thus including the specific aspects of a segment and maintenance event, but also sharing common characteristics. The prior distributions of both univariate and multivariate Wiener modelling approach are set similarly to the section 2.2 above, but the parameters $s_q^{\mu}$ in Eq. (7), $s_q^{\sigma}$ in Eq. (9), and $m_q^z, s_q^z$ in Eq. (11) will follow hyperprior distribution. The hyperparameters of the priors for both multivariate and univariate models were further sampled from hyperprior distributions:

$$s_q^{\mu} \sim \mathcal{U}\left(a_q^{\mu}, b_q^{\mu}\right) \qquad s_q^{\sigma} \sim \mathcal{U}\left(a_q^{\sigma}, b_q^{\sigma}\right)$$

$$m_q^z \sim \mathcal{L}og\mathcal{N}orm\left(M_q^z, S_q^z\right) \qquad s_q^z \sim \mathcal{U}\left(a_q^z, b_q^z\right)$$

The uniform hyperpriors (with large range) used for most of the hyperparameters ensure that the hyperpriors are only weakly informative, as suggested in other works [41]. The hyperprior distributions of $m_q^z$ and $s_q^z$ were set so that most of the prior probability for $z_{q,i,k}^{+}$ was between zero and the indicators critical threshold (which can be found from rail maintenance standards).

The parameter fitting procedure for a segment $i$ is conducted via MCMC program in Stan and Python as follows.

**Step 1**. Provide inputs for fitting model:

- Parameters of hyperpriors: $\eta$ and $a_q^{\mu}, b_q^{\mu}, a_q^{\sigma}, b_q^{\sigma}, M_q^z, S_q^z, a_q^z, b_q^z\ \forall q$
- Data for fitting: time of TRC runs $\{t_1, t_2, \dots, t_K\}$ and track geometry indicators $\widehat{\mathcal{D}}_i = \left\{\mathbf{z}_{i,k} \middle| k = 1,2, \dots, K\right\}$

**Step 2**. Set up the hierarchical Bayesian structure:

- Hyperpriors: $s_q^{\mu} \sim \mathcal{U}\left(a_q^{\mu}, b_q^{\mu}\right)$ ; $s_q^{\sigma} \sim \mathcal{U}\left(a_q^{\sigma}, b_q^{\sigma}\right)$ ; $m_q^z \sim \mathcal{L}og\mathcal{N}orm\left(M_q^z, S_q^z\right)$ ; $s_q^z \sim \mathcal{U}\left(a_q^z, b_q^z\right)$
- Priors: $z_{q,i,k}^{+} \sim \mathcal{LN}\left(m_q^z, s_q^z\right)$ ; $\mathbf{R}_i \sim \mathcal{LKJ}(\eta)$ ; $\sigma_{q,i} \sim \mathcal{H}alf\mathcal{N}orm\left(s_q^{\sigma}\right)$ ; $\mu_{i,q} \sim \mathcal{H}alf\mathcal{N}orm\left(s_q^{\mu}\right)$
- Likelihood: As Eq. (6)

**Step 3**. Obtain posteriors of model parameters $\boldsymbol{\mu}_i, \boldsymbol{\Sigma}_i$ and $\left\{\mathbf{z}_{i,k}^{+} \middle| k \in \mathbb{M}_i\right\}$ via Hamiltonian Monte Carlo algorithm [42].

**Step 4**. Obtain posterior predictive sample: For each set $\{\boldsymbol{\mu}_i, \boldsymbol{\Sigma}_i, \mathbf{z}_{i,k}^{+}\}$ drawn from posterior sample, generate $N_{path}$ degradation paths of track geometry.

# 3 Data description and processing

## 3.1 Data available for this study

The developed multi-indicator degradation model of track geometry is applied to a real case study of suburban train network in Queensland, Australia. For this study, an 18 km-long track, which belongs to this suburban train network, is used for analyses and result illustration. Some key information regarding track inspection and maintenance is reported in the following paragraphs.

*Inspection by TRV*: The investigated track is scheduled for inspection (i.e., TRV running) every 3 months, however, the inspection interval may vary by ± one month due to scheduling challenges. The TRV running collects deviations (in mm) versus designed values for indicators such as: “top” in left rail and right rail (often referred to as longitudinal alignment), left and right “alignment” (often referred to as “horizontal alignment”), twist, gauge, cant, etc... These parameters are recorded with a sampling interval of 0.25 meters. In this study, left top, right top, left alignment and right alignment were considered as elements of $\mathbf{z}_i$.

*Maintenance (i.e. Tamping)*: After the geometry parameters are collected and processed, a tamping activity can be scheduled according to the current track conditions obtained via data and evaluation from experts. In general, track tamping is planned during a specific interval if any of the geometry indicators are found to exceed pre-defined thresholds in the last inspection. The common thresholds are named as M3, M1, D7 and D1 with descriptions as follows:

- M3: tamping should be conducted in next 3 months,
- M1: tamping should be conducted in next 1 month,
- D7: tamping should be conducted in one week, and
- D1: tamping must be conducted immediately.

## 3.2 Track segmentation and maintenance identification

Like other studies on track geometry deterioration, the track is divided into segments of similar length: in this case 182 segments ($N_s = 182$), each of approximately 100 m. Given a sampling interval of 0.25 meters, each segment will therefore contain approximately 400 values for each indicator for each segment, but the degradation model only describes the evolution of one scalar value per geometric indicator. Options to summarize the 400 measurements into a single $z_{q,i,k}$ include using the mean absolute deviation, the standard deviation of the absolute deviation, the maximum or the 95th percentile. The previous study [9] discussed the advantages and drawbacks of these statistics. In this study, the maximum absolute deviation is chosen as the representative statistical index as recommended by the rail operator.

The modelling approach described in Section 2 assumes the knowledge of the inspection intervals $k \in \mathbb{M}_i$ when maintenance actions are performed on each specific segment $i$. In reality, this information is not fully available or is not sufficiently accurate to identify the segment where maintenance was conducted. Therefore, a pre-processing of the inspection data is conducted to determine whether a tamping action was performed between two successive inspections. In particular, a maintenance action is assumed to have been performed in all the inspection intervals that show a simultaneous drop in all geometry indicators.

**Fig. 2** shows an example of the application of the pre-processing technique based on the evolution of TRV measurements for two example segments. Maintenance actions are identified based on the drop of the indicators (i.e. Top Left (10), Top Right (10), Top Left (6), Top Right (6), L-Alignment (10), R-Alignment (10) [1]) in the intervals 23-Jan to 12 Mar 2020 (left) and 12-Mar to 25 Jun 2020 (right). A manual analysis of work orders confirms that tamping occurred on these segments on 1 Feb 2020 and 15 Jun 2020, respectively.

This approach is expected to deliver reasonable results in most cases, yet a small level of error is expected. In fact, some inconsistences with the work-order database have been identified. **Fig. 3** shows two examples of these inconsistences. The situation on the left indicates that there is a tamping event for this track segment recorded in the work-orders on 7 April 2020, but the geometry-based identification cannot confirm this maintenance, given no improvement in the geometry indicators. An example of the opposite case is shown in **Fig. 3** (right), where a tamping is identified based on the drop of all geometric deviations but does not show in the work-order records. It is likely that these inconsistencies are largely due to the insufficient location specificity of the work-order records, but it is possible that some ineffective tampings would be missed and some (very rare) simultaneous improvements in indicators could result in false positives. Nonetheless, after consultation with experts, the objective analysis of the statistical maintenance identification was preferred given the challenges in the work order timing and space specification.

[1] Values 10 and 6 in bracket relates to the chord length (in meters) of measurement system.

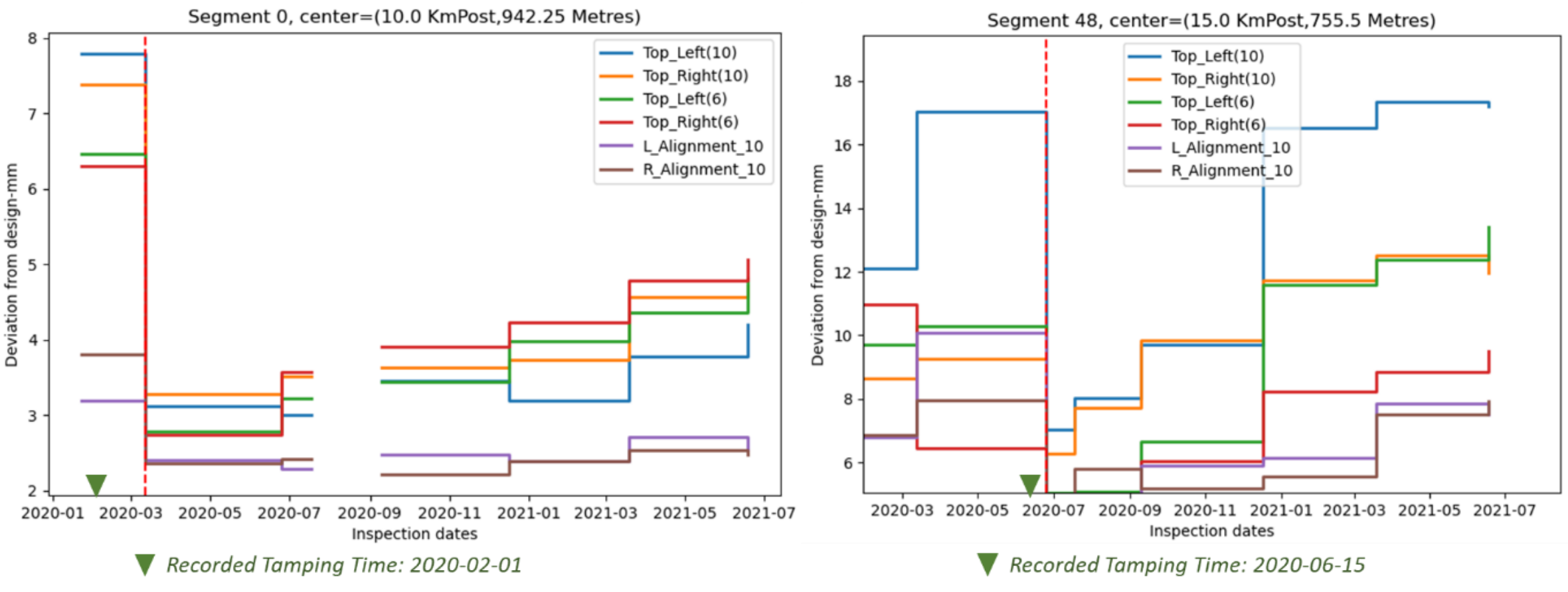

**Fig. 2** Illustration of maintenance identification - consistence with workorder

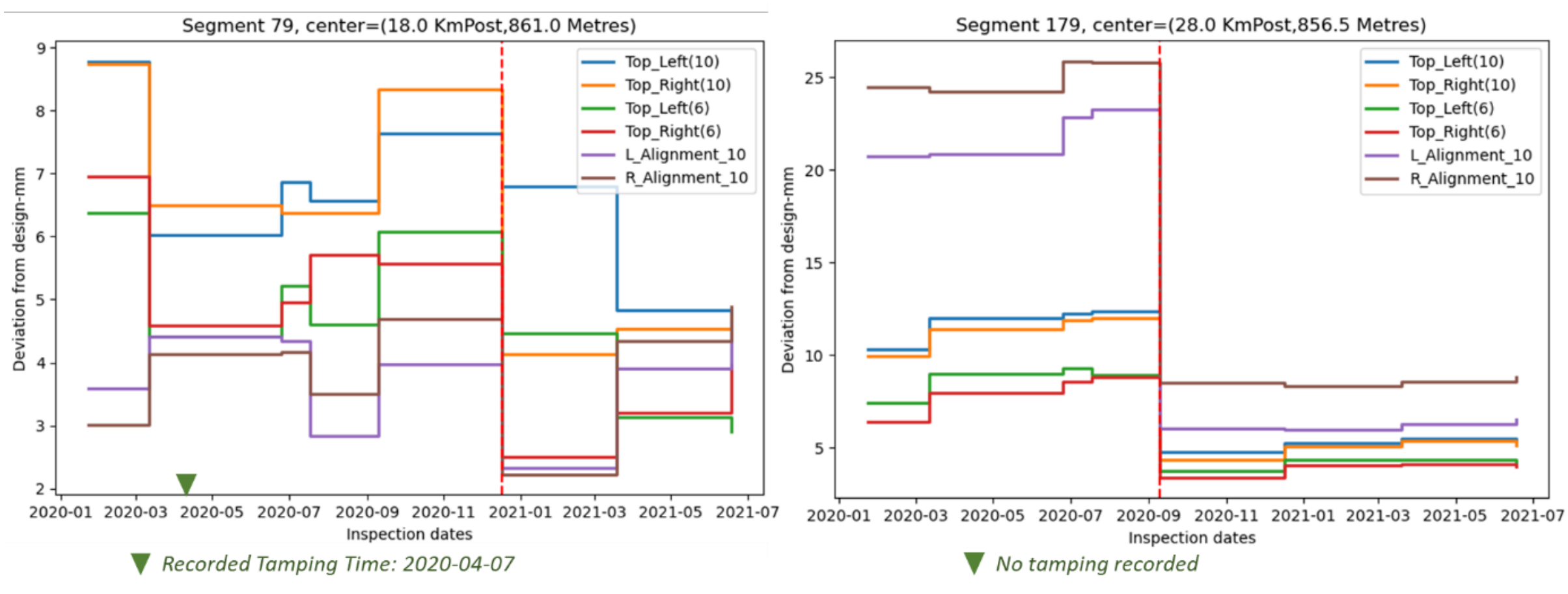

**Fig. 3** Illustration of maintenance identification - inconsistence with workorder

## 4 Results of Degradation models

This section presents the results of the newly proposed multi-variate track degradation model and its benchmark with traditional univariate models. The indicators Top Left (6), Top Right (6), Left Alignment (10) and Right Alignment (10) are considered. The models are fitted according to the Bayesian and hierarchical Bayesian approaches as discussed in section 2 with parameters of priors and hyperprior distributions as shown in Table 1 and Table 2, respectively.

Table 1: Prior parameters for individual segment fitting

| $\boldsymbol{s_q^{\mu} = 2}$ | $\boldsymbol{m_q^{z} = 2.3}$ |
|---|---|
| $\boldsymbol{s_q^{\sigma} = 5}$ | $s_q^{z} = 2$ |
| | $\eta = 1$ |

Table 2: Hyperpriors setting for hierarchical Bayesian fitting

| **Univariate Bayesian Model** | **Multivariate Bayesian Model** |
|---|---|
| $\boldsymbol{s_q^{\mu} \sim \mathcal{U}(0\,,10)}$ | $s_q^{\mu} \sim \mathcal{U}(0\,,10)$ |
| $\boldsymbol{s_q^{\sigma} \sim \mathcal{U}(0\,,10)}$ | $s_q^{\sigma} \sim \mathcal{U}(0\,,10)$ |
| $\boldsymbol{m_q^{z} \sim \mathcal{L}og\mathcal{N}orm(2.3\,,1)}$ | $m_q^{z} \sim \mathcal{L}og\mathcal{N}orm(2.3\,,1)$ |
| $\boldsymbol{s_q^{z} \sim \mathcal{U}(0\,,2)}$ | $s_q^{z} \sim \mathcal{U}(0\,,2)$ |
| | $\eta = 1$ |

In practice, hyperparameters or prior parameters are selected based on the expertise of track geometry maintenance teams from rail operators. If expert advice is unavailable, these parameters are chosen to meet two criteria: (1) ensuring the prior range is sufficiently broad (i.e., only weakly informative) and (2) maintaining a high probability that random values drawn from the prior fall within a reasonable range. For example, in this study, the $\mu$ parameters of $m_q^z$ is selected at 2.3 that can make the (prior) median of $z_{q,i,k}^+$ around 10.

## 4.1 Hierarchical Bayesian fitting of the Multivariate Wiener model

Thanks to the hierarchical Bayesian approach, the posterior distribution of each $\mathbf{z}_{i,k}^+$ (geometry indicators after maintenance) can be estimated considering data from all segments (index $i$) and all events (index $k$). The posterior distributions of $\mathbf{z}_{i,k}^+$ are shown in **Fig. 4** as boxplots for all identified maintenance events. Note that each maintenance event posterior is shown independently without regard to the time and segment in which the event occurred. Each subplot shows the post-maintenance value of a particular component $z_q^+$ after one of the 73 identified maintenance events. The variation in the effect of tamping can be clearly seen is also indicated via PDFs functions of $z_q^+$ presented in **Fig. 5**, which show the predictive distributions of $\mathbf{z}^+$ (found via sampling from the hyperparameter distributions), which characterize the randomness of post-maintenance values for the different indicators. The distributions of the two Top indicators are similar, as are those of the alignment indicators.

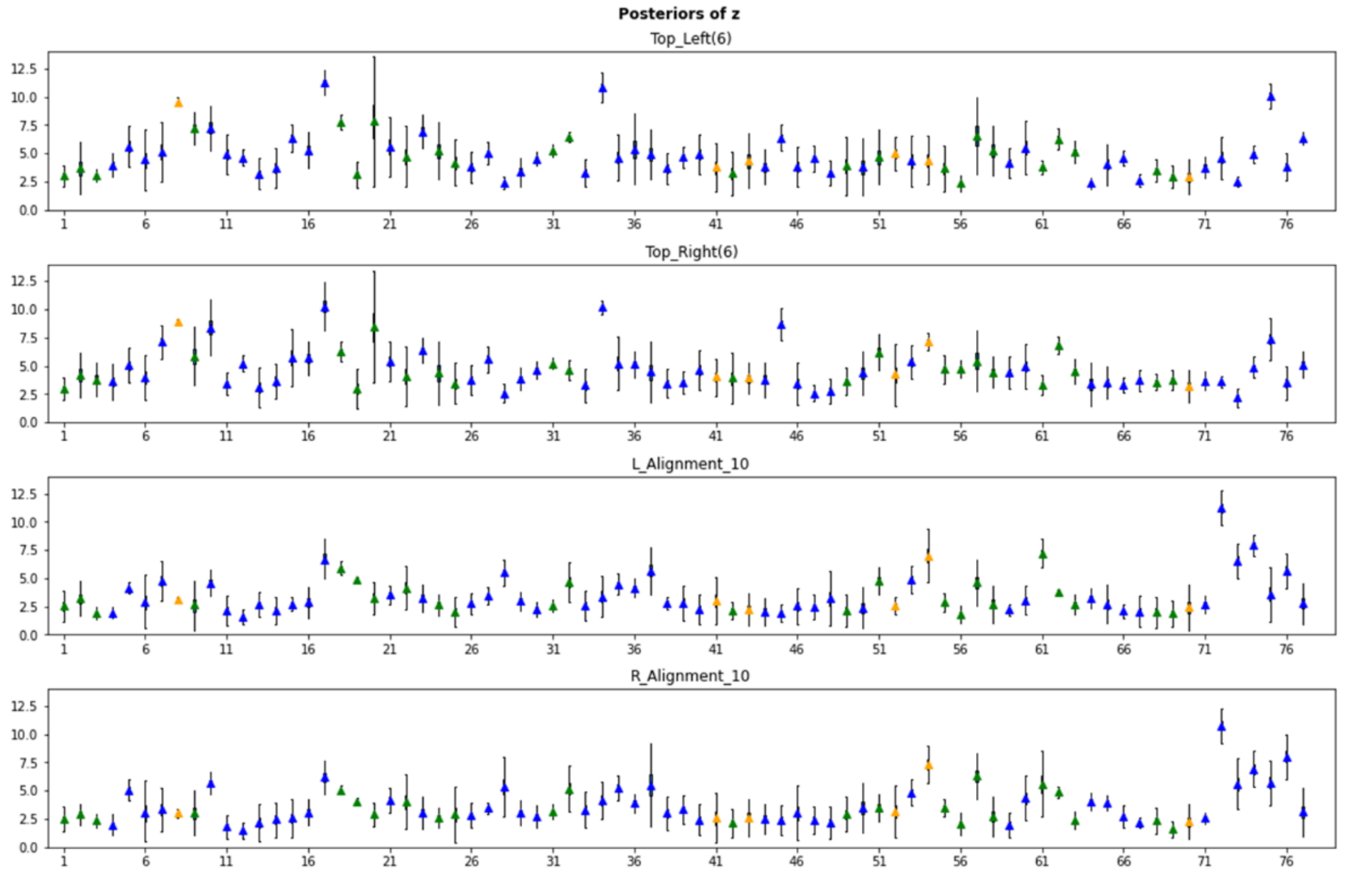

**Fig. 4** Posteriors of $\mathbf{z}_{i,k}^{+}$

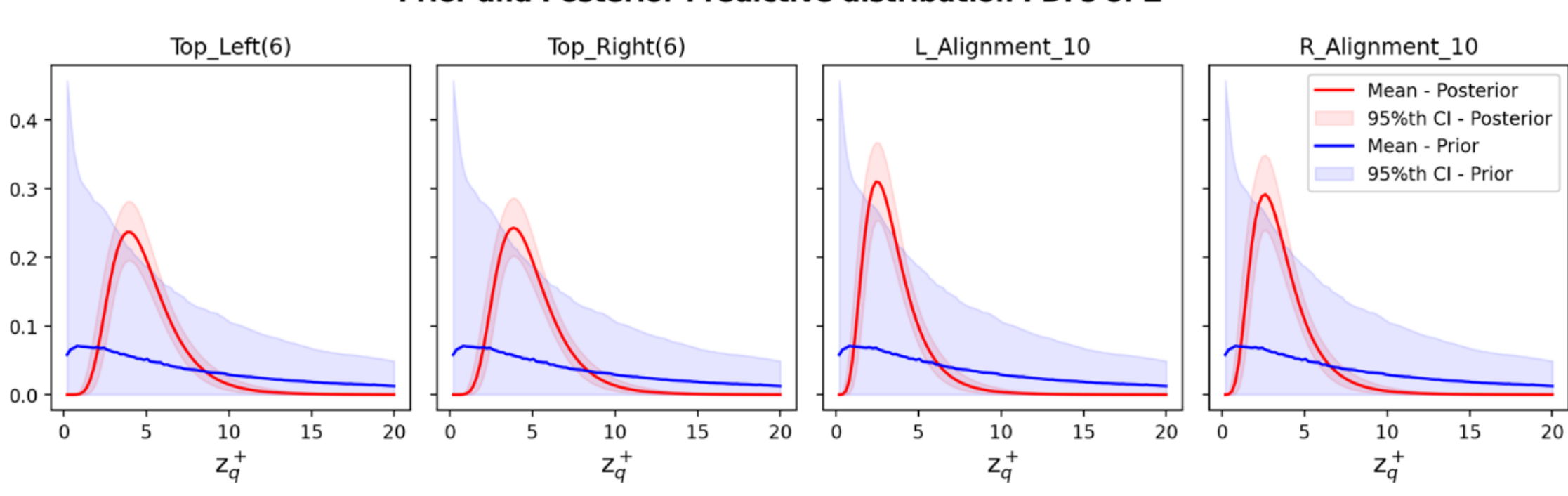

**Fig. 5** Marginal Prior and Posterior Predictive Distributions of $\mathbf{z}^{+}$ components

**Fig. 6** and **Fig. 7** show the posterior distributions of vector $\boldsymbol{\mu}_i$ and the square root of the diagonal terms of $\boldsymbol{\Sigma}_i$ (marginal standard deviations) for each segment $i$ (x-axis). The drift $\boldsymbol{\mu}_i$ is strikingly similar across segments except for a few high-degradation segments. Again, the similarities of $\mu_q$ and $\sigma_q$ occur for two pair left and right indicators of Top and Alignment. It means that there is correlation between 2 longitudinal indicators and 2 alignment indicators. This correlation will be clarified in estimation of correlation matrix $\boldsymbol{\Sigma}_i$. While the difference of $\boldsymbol{\mu}_i$ across segments is small, the variation of the marginal standard deviations is much larger (see **Fig. 7**). It implies that the difference of degradation speed is mostly driven by $\boldsymbol{\sigma}_i$, a somewhat negative (yet novel) finding for maintenance planning, given that randomness has a higher impact than drift. Some anecdotal evidence of this is the fact that segment 42 which has the highest standard deviation value is the fastest degradation segment in this case study.

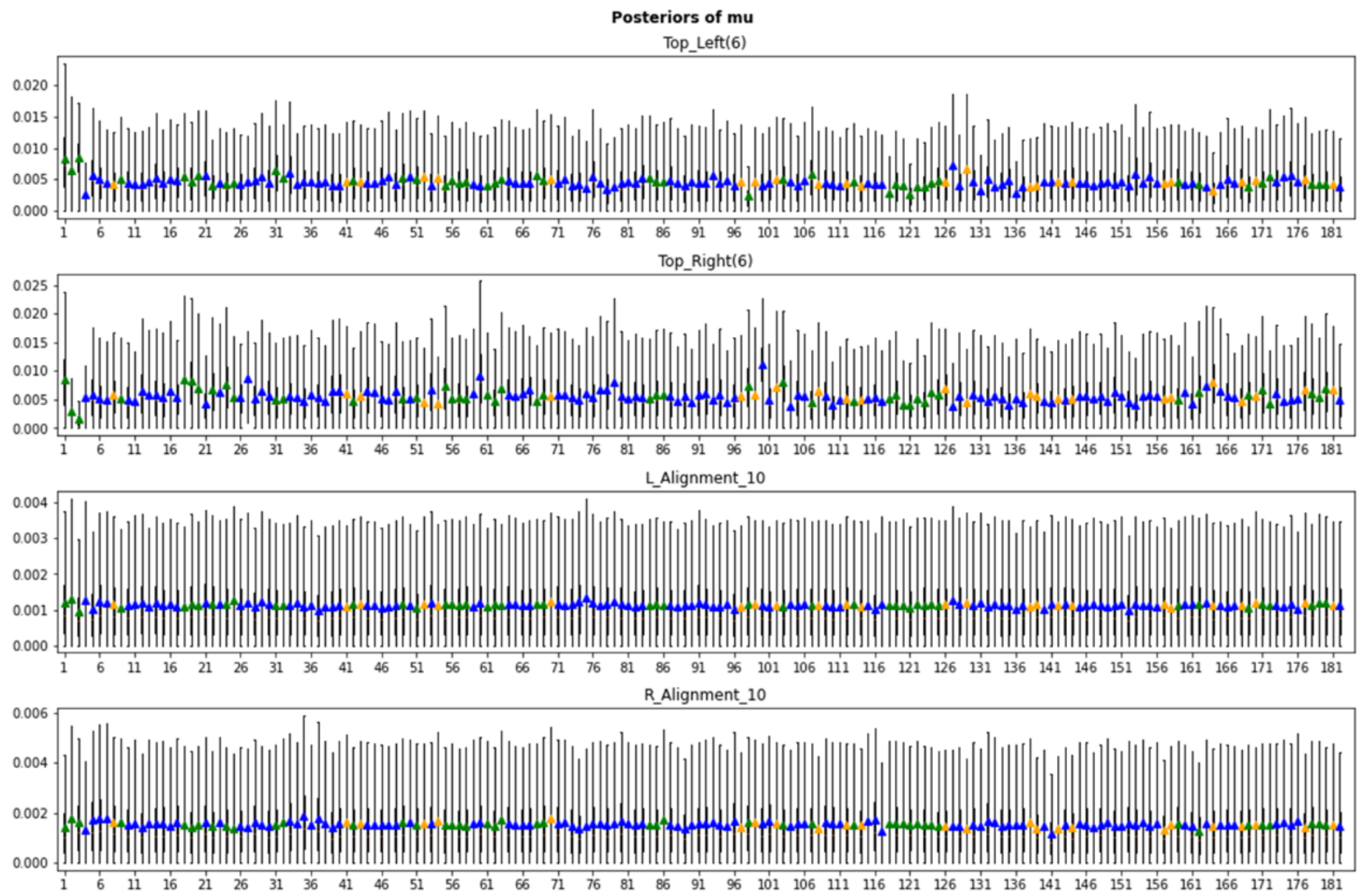

**Fig. 6** Posterior distributions of $\mu_{i,q}$ for each segment, multi-variate model

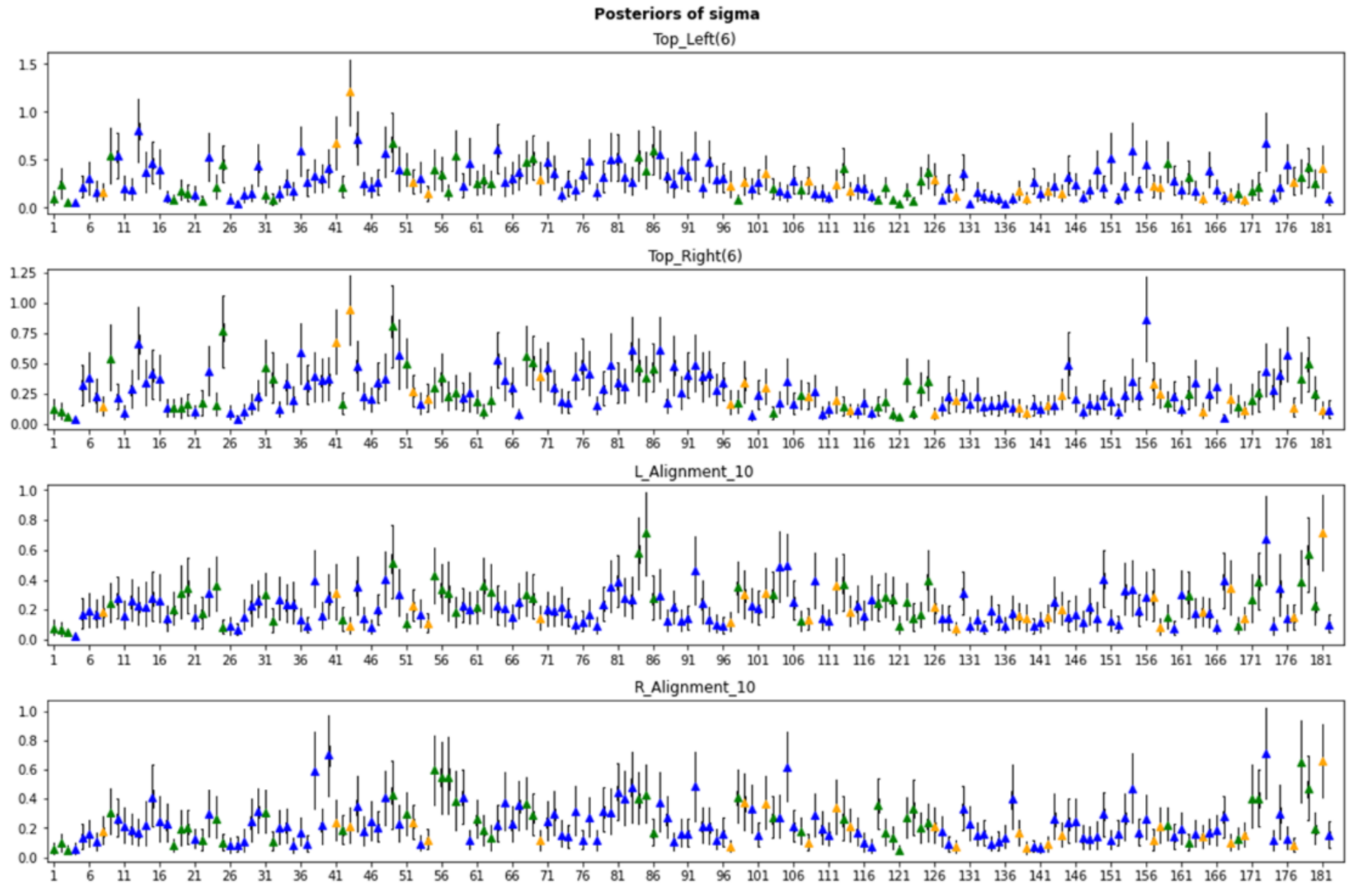

**Fig. 7** Posterior distributions of $\sigma_{i,q}$ for each segment (diagonal elements of $\mathbf{\Sigma}_i$), multi-variate model

**Fig. 8** shows the posterior distribution of the elements of the correlation matrix $\mathbf{R}_i = [r_{ij}]_{i,j=0..3}$ over all segments. Please note that the diagonal terms of this matrix would be one, so the plot instead shows the histograms of $\sigma_i$. Since the prior distribution of correlation (i.e., LKJ distribution) is set with $\eta = 1$, it means that no correlation among all indicators is supposed *a priori*. However, the posteriors shown in the figure indicate clearly that data supports the presence of correlations between the two Top indicators (i.e., $r_{10} \approx 0.8$) and two Alignment indicators (i.e, $r_{32} \approx 0.6$). On the contrary, no clear evidence of correlation is visible between vertical and horizontal indicators with $r_{20}, r_{30}, r_{21}, r_{31}$ around $[0\,, 0.2]$.

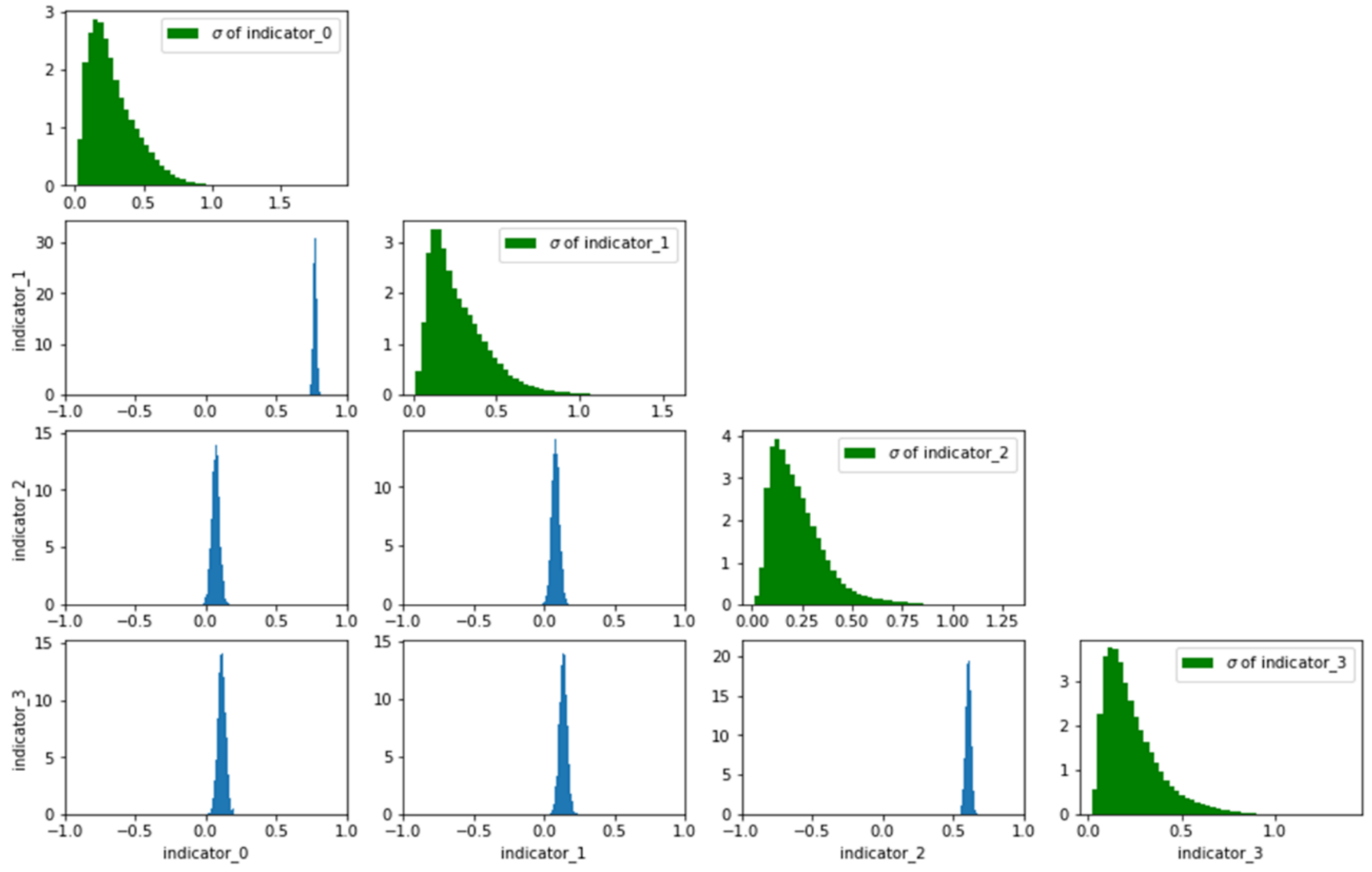

**Fig. 8** Posterior distribution for $\mathbf{R}_i$ for all segments. Diagonal elements indicate posterior distributions of $\sigma_i$ over all segments.

## 4.2 Comparison between hierarchical and individual Bayesian fitting

The posteriors of $\mathbf{z}_{i,k}^{+}$, $\mu_{i,q}$ and $\sigma_{i,q}$ obtained through individual Bayesian fitting are shown in **Fig. 9**, **Fig. 10** and **Fig. 11**, respectively. In general, the pattern across segments or maintenance events (for $\mathbf{z}_{i,k}^{+}$) in the posteriors from individual fitting are similar to those from the hierarchical fitting shown earlier. However, the posteriors from individual fitting exhibit more variability compared to those from hierarchical fitting. Additionally, the range of posteriors from individual fitting is wider than that from hierarchical fitting in drift parameter $\mu$ (with credible interval $[0\,, 0.118]$ $vs.$ $[0\,, 0.01]$) and deviation $\sigma$ ($[0.03\,, 0.832]$ $vs.$ $[0.032\,, 0.558]$) (see **Fig. 12**). This is because the data sample used for individual fitting of each segment is smaller, whereas hierarchical fitting utilizes data from all segments, contributing to a more robust fitting process. This highlights the benefits of the hierarchical Bayesian approach, particularly when the dataset is small.

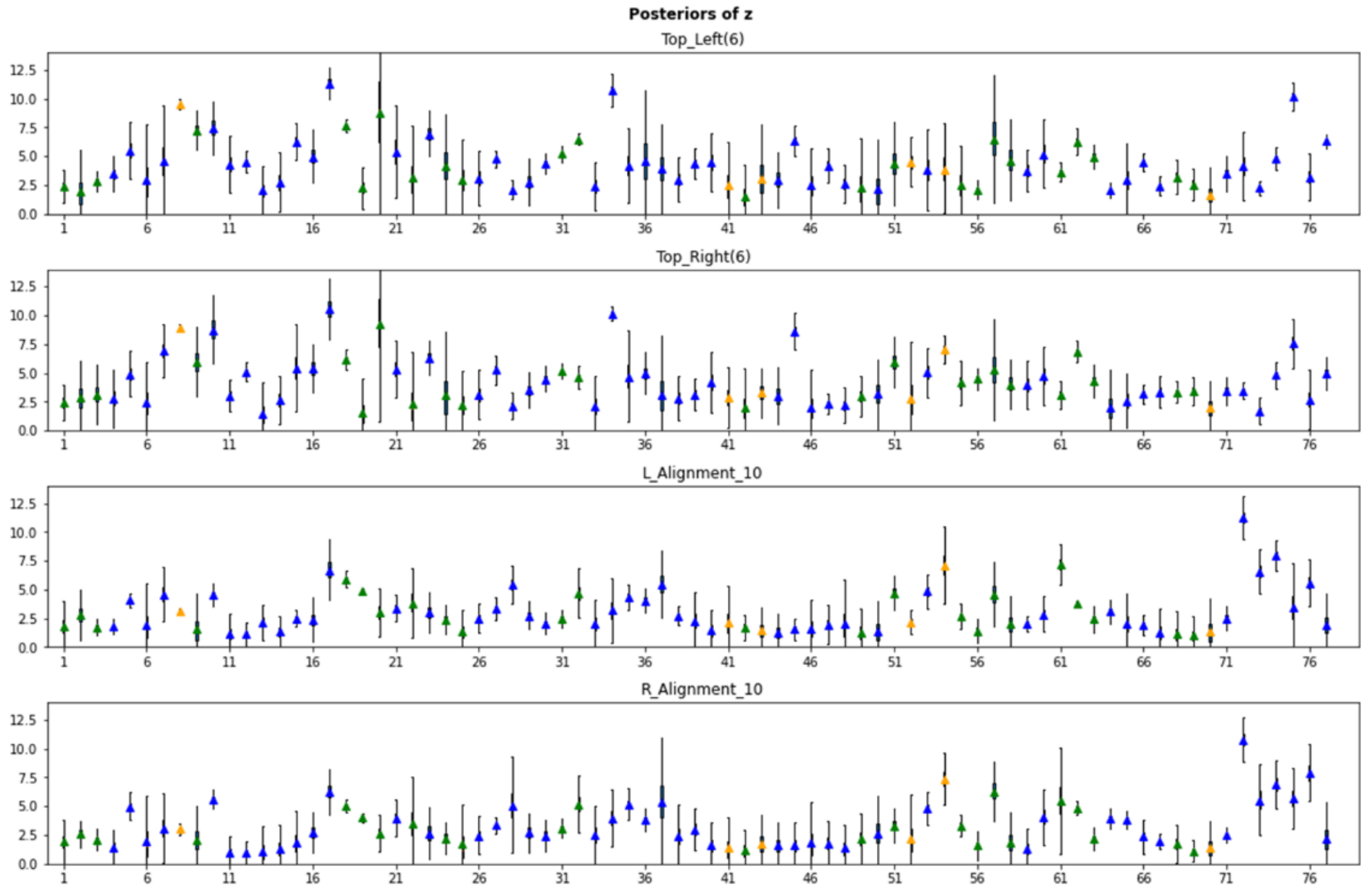

**Fig. 9** Individual fitting: Posteriors of $\mathbf{z}_{i,k}^{+}$

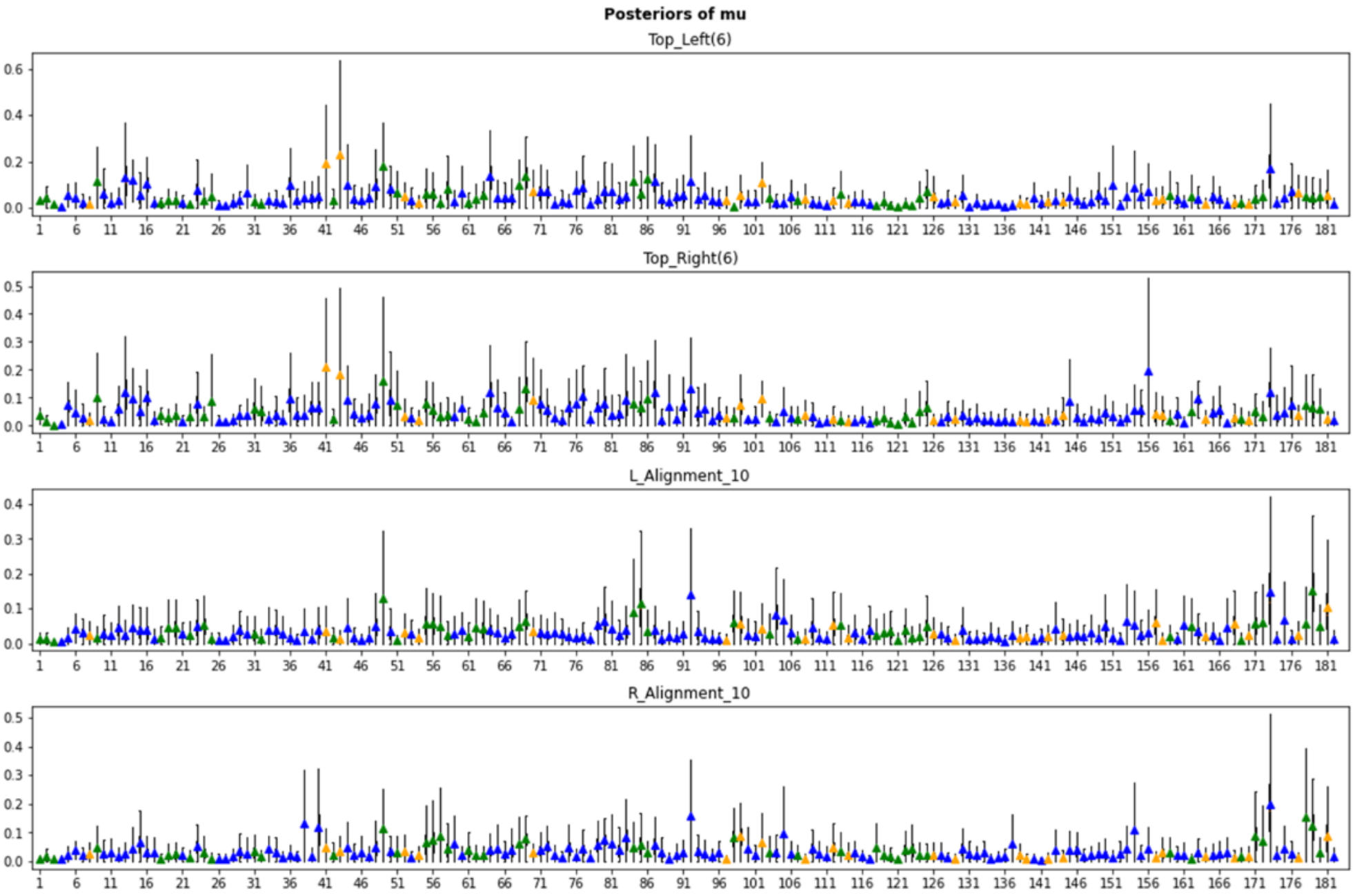

**Fig. 10** Individual fitting: Posterior distributions of $\mu_{i,q}$

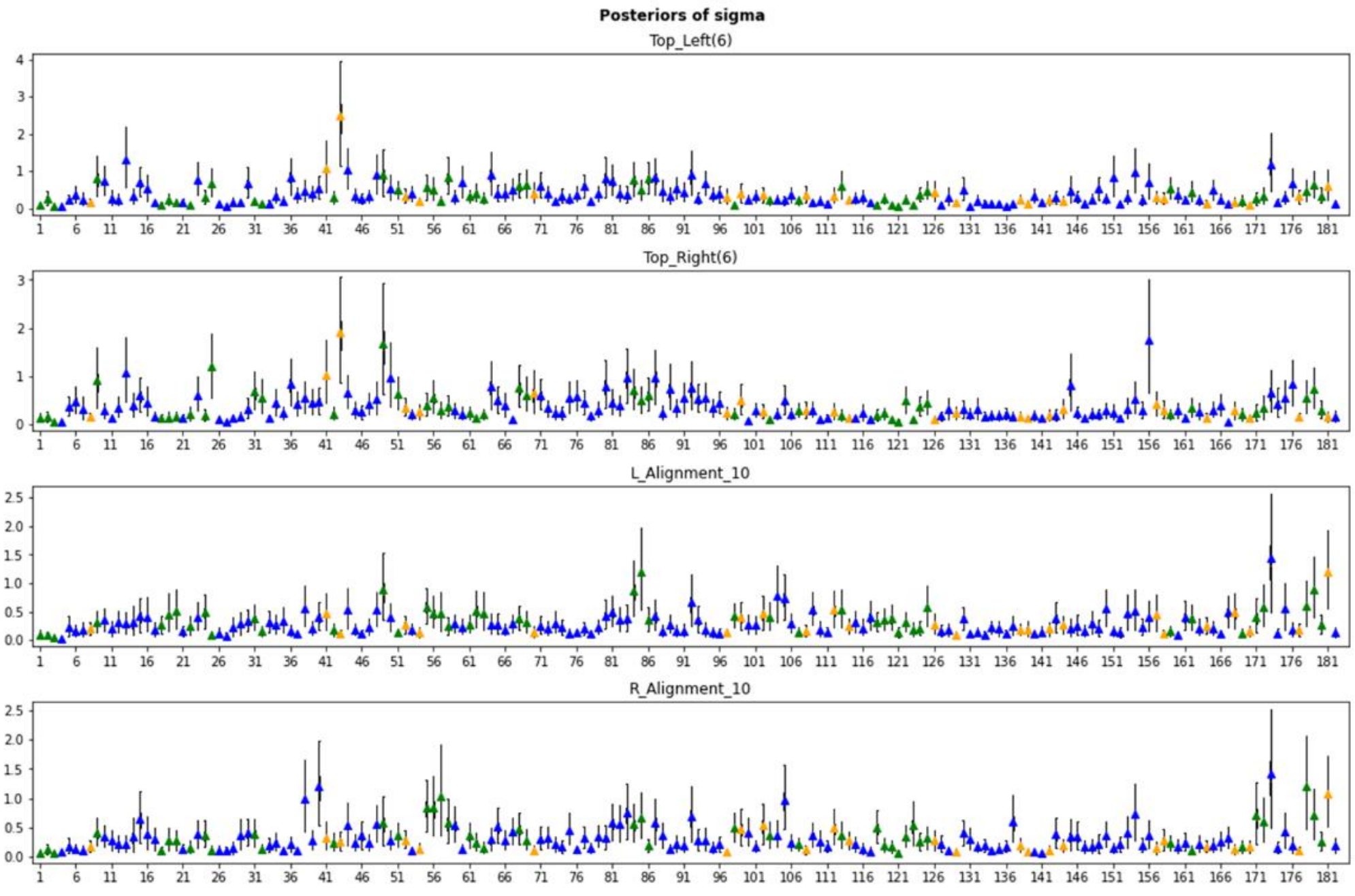

**Fig. 11** Individual fitting: Posterior distributions of $\sigma_{i,q}$

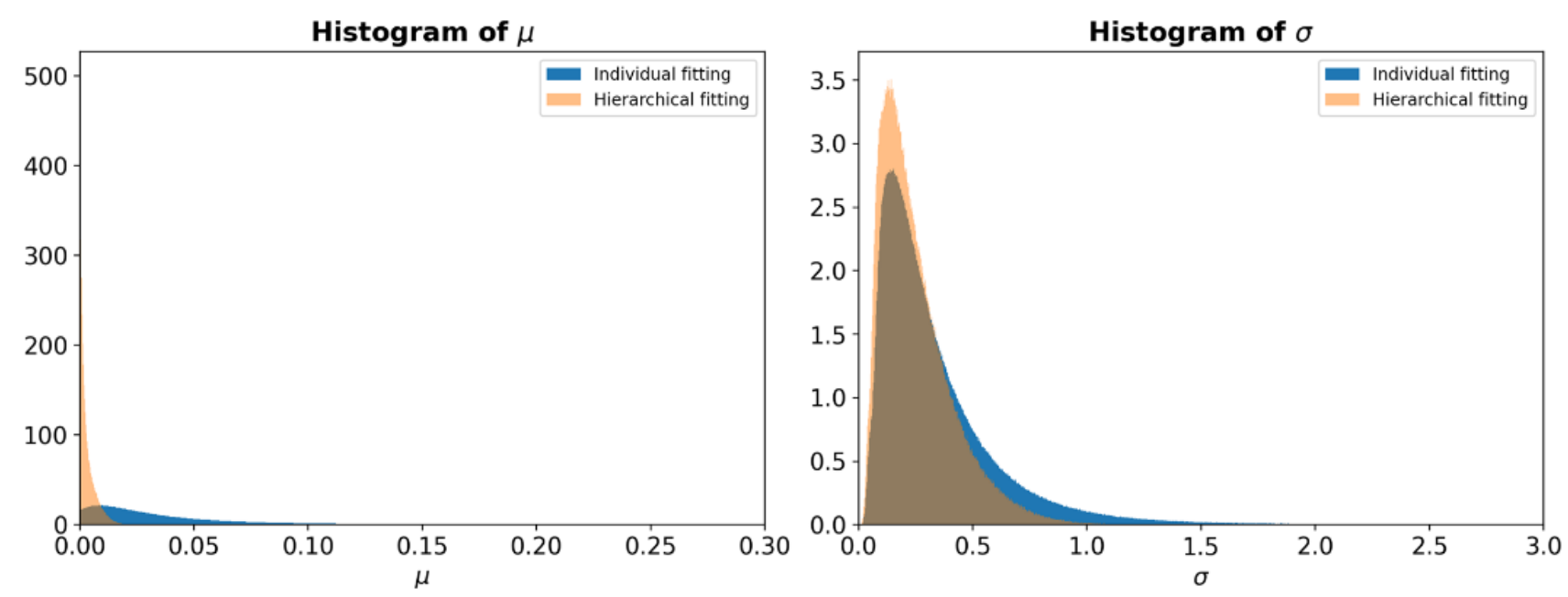

**Fig. 12** Histograms of posteriors of $\mu$ and $\sigma$ from two fitting approaches

## 4.3 Validation of the model's predictive capabilities

The effectiveness of the model in predicting degradation was evaluated by dividing the inspection database into training and validation sets. The last three inspection measurements were used for validation, while the rest were used for fitting. Fitting and prediction results using the multivariate model are shown in **Fig. 13** and **Fig. 14** for cases with and without maintenance. The model fits the data well, including trend corrections in the case of maintenance. All measured data points are within the 95% credible interval of the model, both with and without maintenance. For the entire track of 182 segments, 88.9% of testing data points fall within the credible zone, while 5.3% fall above this zone (indicating a prediction that misses a defect).

These results show that the proposed multivariate Wiener model is suitable for predicting and analyzing track geometry degradation, with longitudinal indicators (i.e., Top left and right) showing more variation and degradation than Alignment indicators. This suggests that longitudinal indicators are likely the most common triggers for maintenance actions — a topic to be explored in the next section.

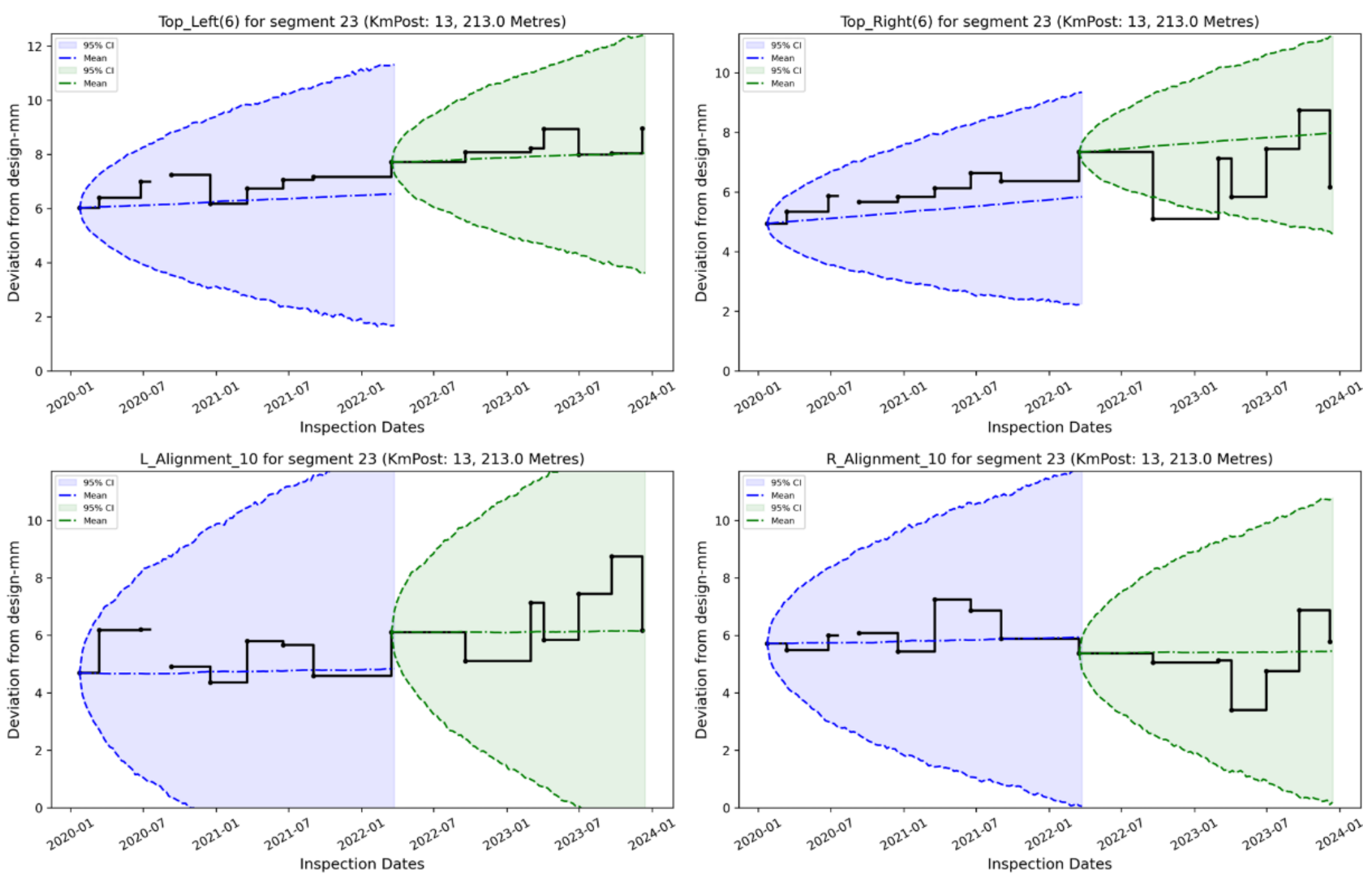

**Fig. 13** Multivariate Model: Fitting and prediction in case of no maintenance

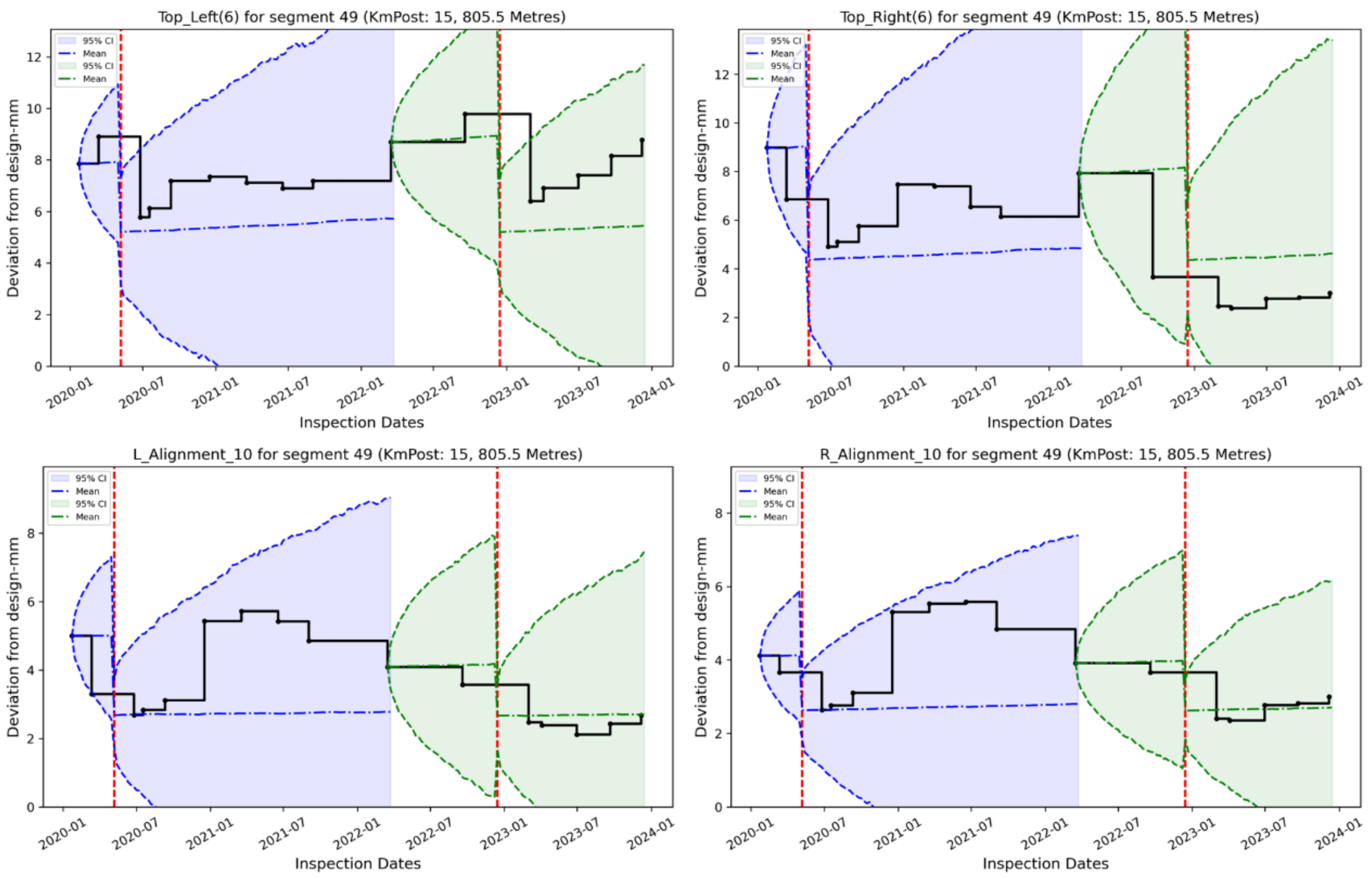

**Fig. 14** Multivariate Model: Fitting and prediction in case of maintenance occurrence

For the univariate Wiener model, **Fig. 15** and **Fig. 16** show the fitting and prediction results for two scenarios: with and without maintenance. Both scenarios exhibit similar performance for individual indicators, but the multivariate model is slightly better, with fewer points outside the credible interval (compare between **Fig. 13** vs. **Fig. 15**). Across the entire track, the univariate model has poorer predictive performance: 63.8% of the test data falls within the credible zone, while 8% exceeds it. The multivariate model also produces faster-growing prediction intervals, offering less conservative estimates of when an indicator will reach the maintenance threshold. However, its key advantage lies in capturing correlations between indicators, which is not evident from individual plots. This benefit will become clearer in the next section, which analyzes the first hitting time of any indicator.

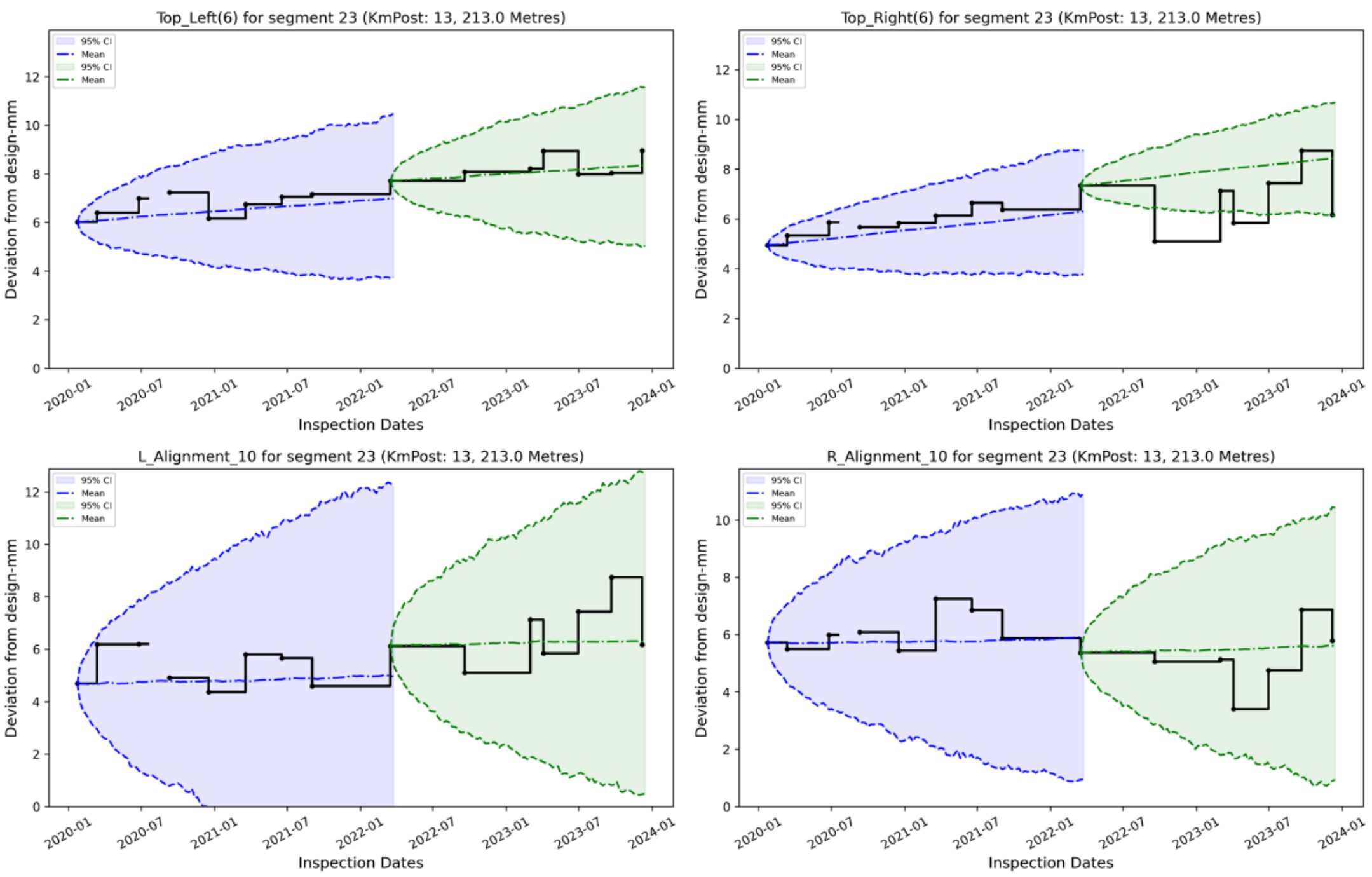

**Fig. 15** Univariate Model: Fitting and prediction in case of no maintenance

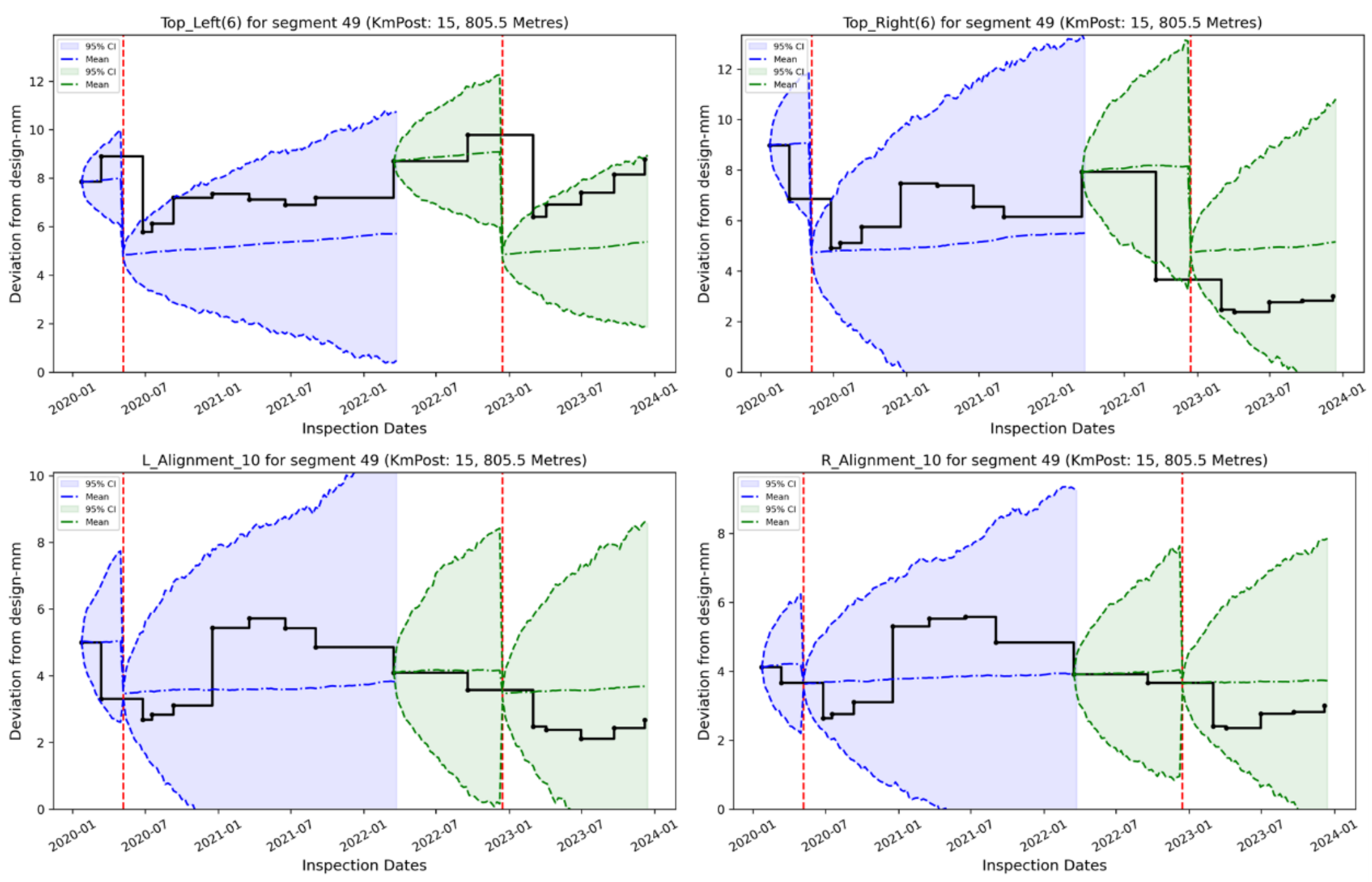

**Fig. 16** Univariate Model: Fitting and prediction in case of maintenance occurrence

## 4.4 Analysis of the first hitting time of any indicator

A benefit of the proposed track geometry degradation model is its ability to predict the first time any one indicator exceeds its predetermined maintenance threshold. This first hitting time estimation is based on Monte-Carlo simulations using the degradation model. The behavior of the newly proposed multivariate model can be compared to the alternative of using traditional univariate models. In this section, the hitting times are examined, which are defined by

$$t_{hitting} = \left\{ \min \tau \;\middle|\; \min_{q} \left[ z_{q,th} - Z_{q,i}(\tau) \right] \leq 0 \right\} \tag{12}$$

where $z_{q,th}$ is the threshold for indicator $q$. In other words, the hitting time is defined as the first time where *one* of the $q$ indicators exceeds its threshold. For the multivariate case, $\mathbf{Z}_i(t)$ is predicted by the multivariate Wiener model, while the traditional univariable models will predict the components $Z_{q,i}(t)$ individually. Since in the univariate case does not acknowledge correlation among the indicators, the hitting time described by Eq. (12) will be smaller.

**Fig. 17** presents the distribution of hitting time defined by Eq. (12) according to four maintenance thresholds described above. The x-axis is the hitting time (in weeks) of segment. The hitting time distributions of the multivariate Wiener model are flatter compared with the outputs of the univariate model and show that indeed the univariate model will hit maintenance thresholds sooner. The multivariate model is instead capturing the strong correlation between some pairs of indicators, resulting in behavior that is closer to “worst between two” rather than the “worst among four” scenario.

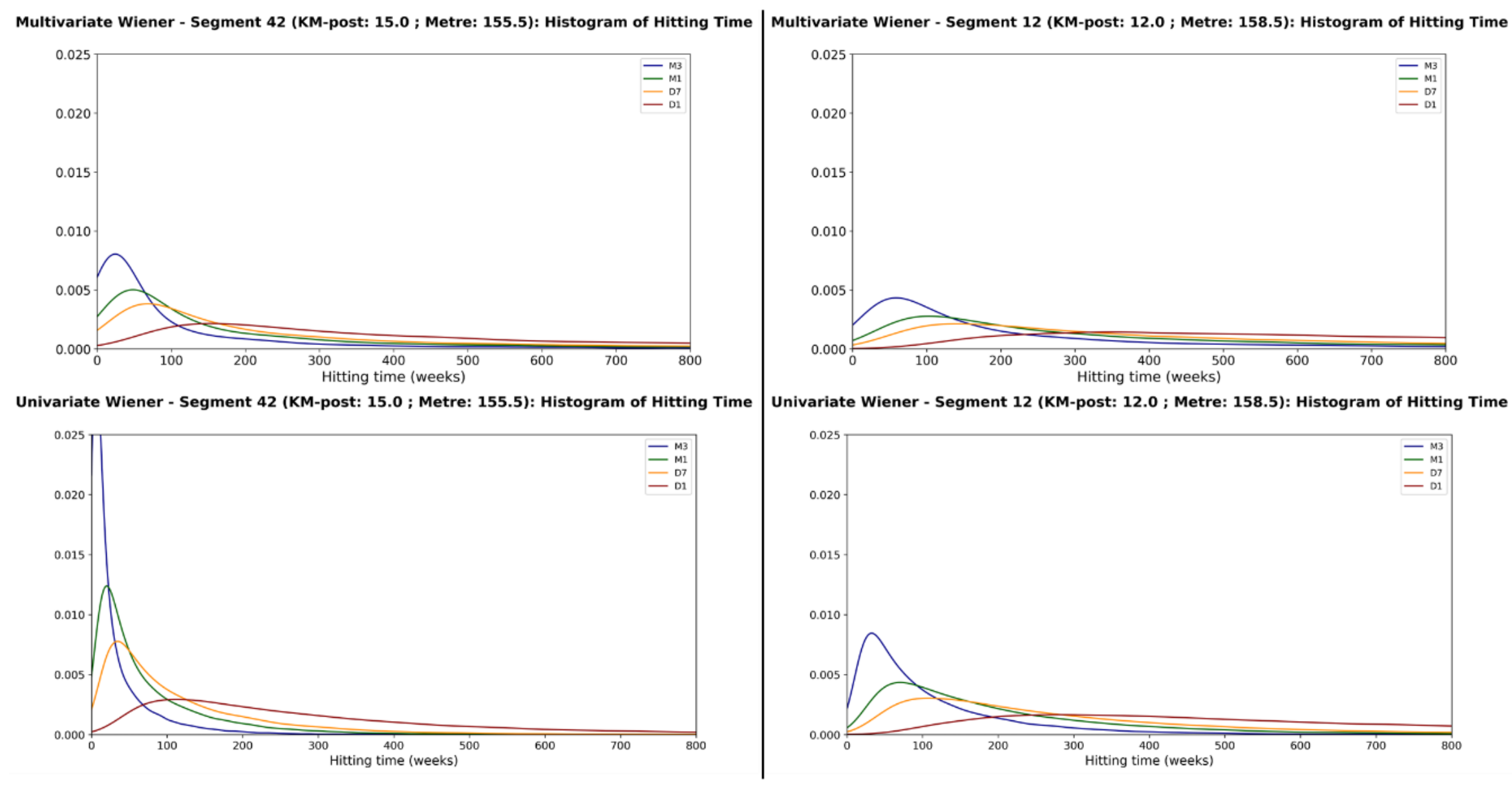

**Fig. 17** Hitting time distributions: Multivariate (upper) and Univariate (lower)

**Fig. 18** shows the probabilities of each individual indicator being the first to hit its threshold (i.e. it is the indicator responsible for the hitting time). It can be seen that the Top indicators are the major causes for triggering maintenance decision with roughly more than 80% of cases. This output also supports the common statement in literature of track geometry research that longitudinal indicators are the most important geometry parameter for maintenance.

**Univariate Wiener - Segment 172 (KM-post: 28.0 ; Metre: 156.5): Hitting Indicator Chances**

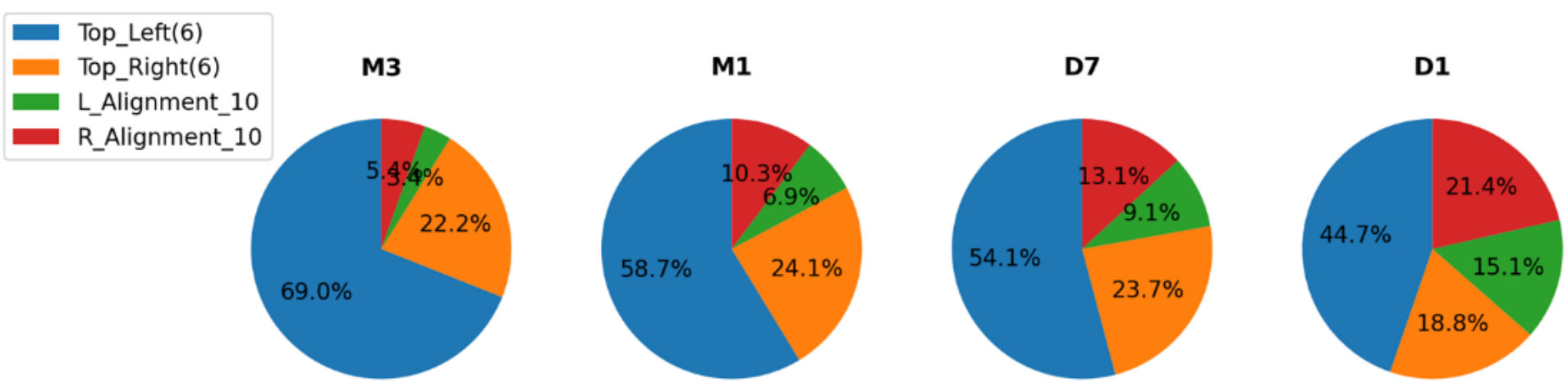

**Multivariate Wiener - Segment 172 (KM-post: 28.0 ; Metre: 156.5): Hitting Indicator Chances**

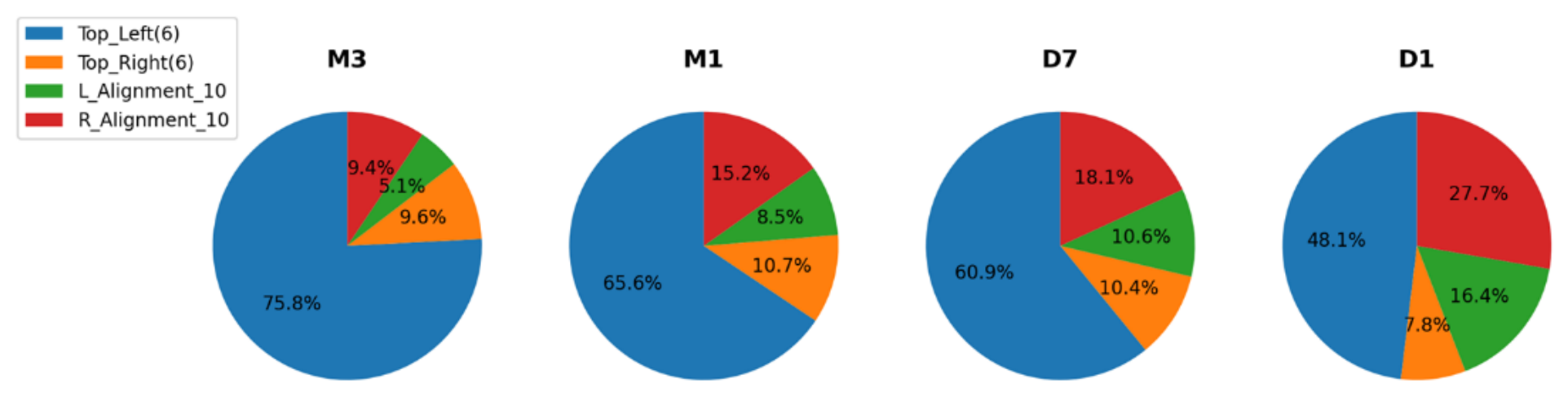

**Fig. 18** Chances of hitting indicators: Multivariate (lower) and Univariate (upper)

## 4.5 Alternative degradation models

### 4.5.1 Modelling combined left and right indicators

This section examines simplifying the track geometry degradation model by reducing the number of indicators. The strong correlation between the left and right sides of Top and Alignment (**Fig. 8**) suggests that using only two indicators—Maximum Top and Maximum Alignment—may maintain accuracy. These composite values represent the worst-case scenario by taking the higher value from each side.

A multivariate model using these two indicators (Max Top and Max Alignment) was developed. Results (**Fig. 19**) show that hitting time distributions remain unchanged compared to the original four-indicator model, indicating no loss of accuracy. However, reducing indicators in univariate models leads to longer predicted hitting times, as fewer independent degradation variables decrease the probability of reaching critical thresholds. Still, the 2-indicator univariate model predicts earlier degradation than multivariate approaches, suggesting that the correlation between Top and Alignment, while modest, significantly impacts predictions.

In summary, track maintenance decision-makers can integrate left and right indicators for each dimension (e.g., Top or Alignment) without compromising accuracy. Additionally, this analysis suggests that traditional single-indicator-based maintenance strategies may be overly conservative, particularly when key track geometry indicators are correlated.

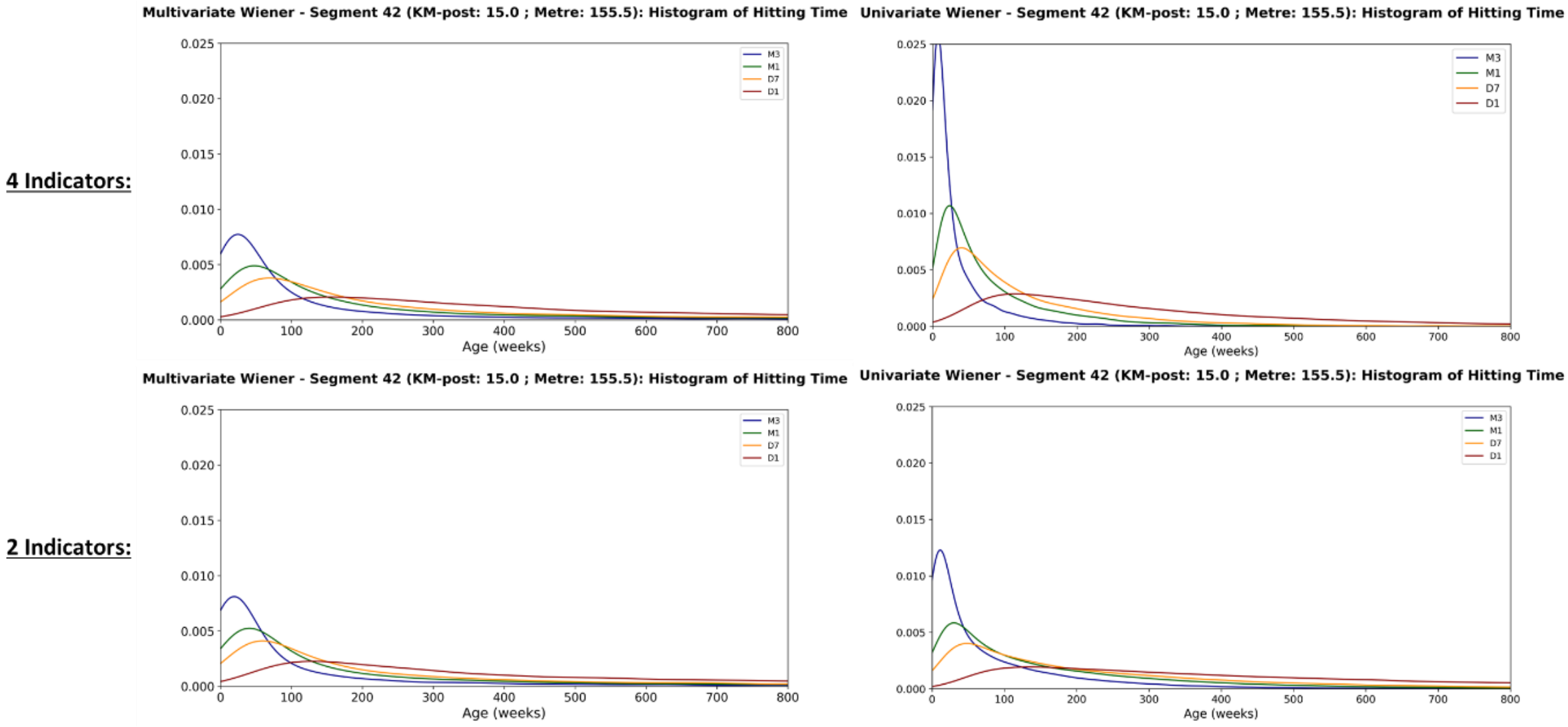

**Fig. 19** 4-indictor model vs. 2-indicator model: Histogram of Hitting Time

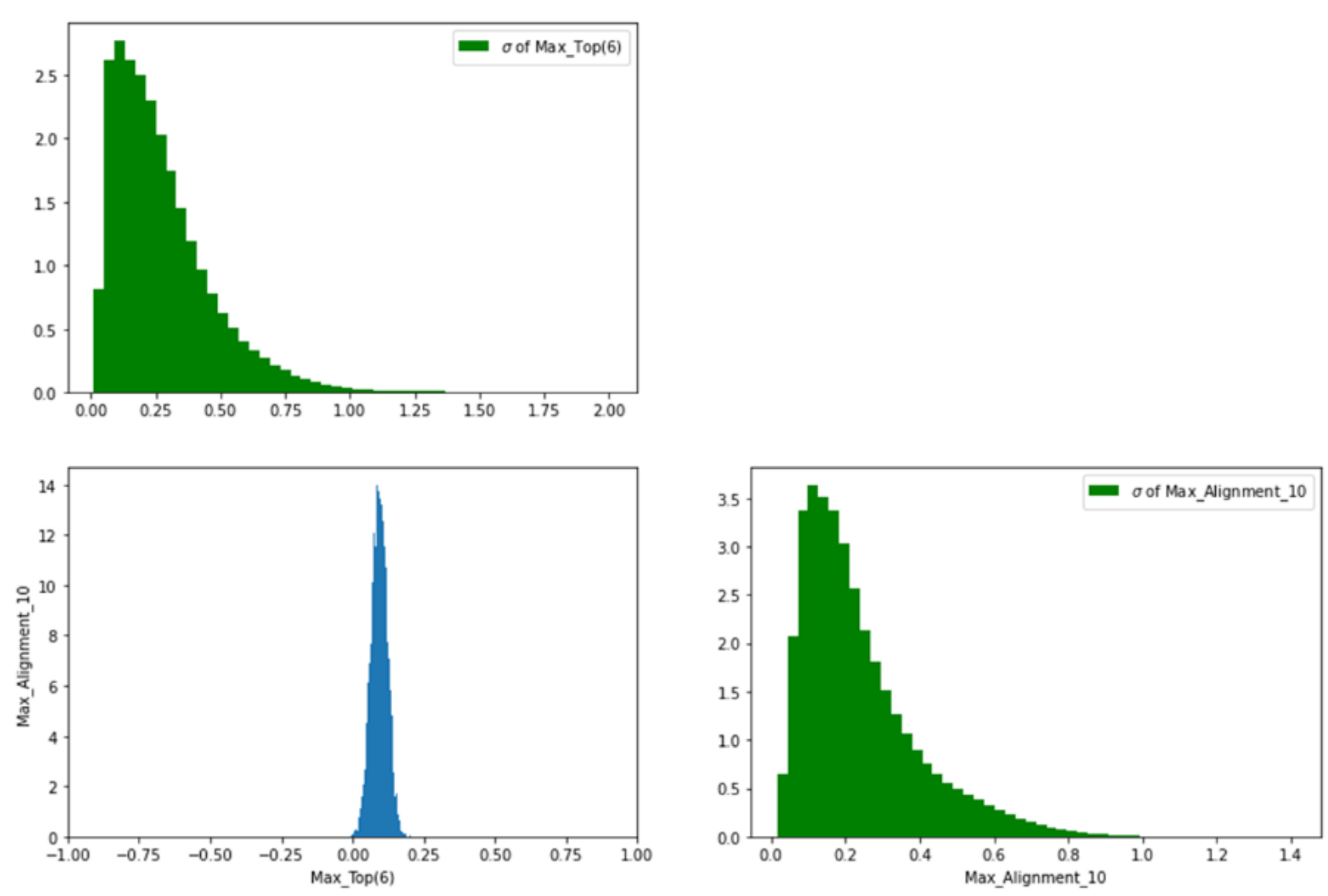

**Fig. 20** 2-indicator model: Correlation

### 4.5.2 Modelling degradation of track geometry indicators in log-space

While linear degradation is a reasonable assumption, one issue with proposed model in Section 2 is that the geometry is that the degradation distributions have support over the entire real line $(-\infty, \infty)$. Yet, track geometry indicators cannot be negative. To address this, the degradation of track geometry can be modeled using the logarithm of the track geometry indicators rather than their direct values. The prediction in log-space will result in an exponential increment of degradation mean which is difficult to justify *a priori*. Therefore, the predictions of the analytically convenient log space model will be compared to those of the nominal model. This section presents a degradation model developed in the log-space of track geometry indicators, incorporating both left and right indicators. The modeling procedure follows the same approach described in Section 2, but with the input data being the natural logarithm of data obtained from the TRC.

**Fig. 21** illustrates the validation of the model developed in the log-space of track geometry indicators. The model remains valid in log-space, as the predicted data fall within the 95th percentile credible interval, both with and without maintenance events. Notably, in the log-space model, the credible intervals are all greater than zero, aligning with the reality that track geometry indicators cannot be negative. The credible intervals for model fitting and prediction are skewed to the right, reflecting the impact of modeling in log-space. Considering the entire track, 82.9% of testing data belongs to the credible zone and 5.3% is higher than the upper limit. This validation provides evidence that the developed log-space model is acceptably accurate for predicting track geometry degradation.

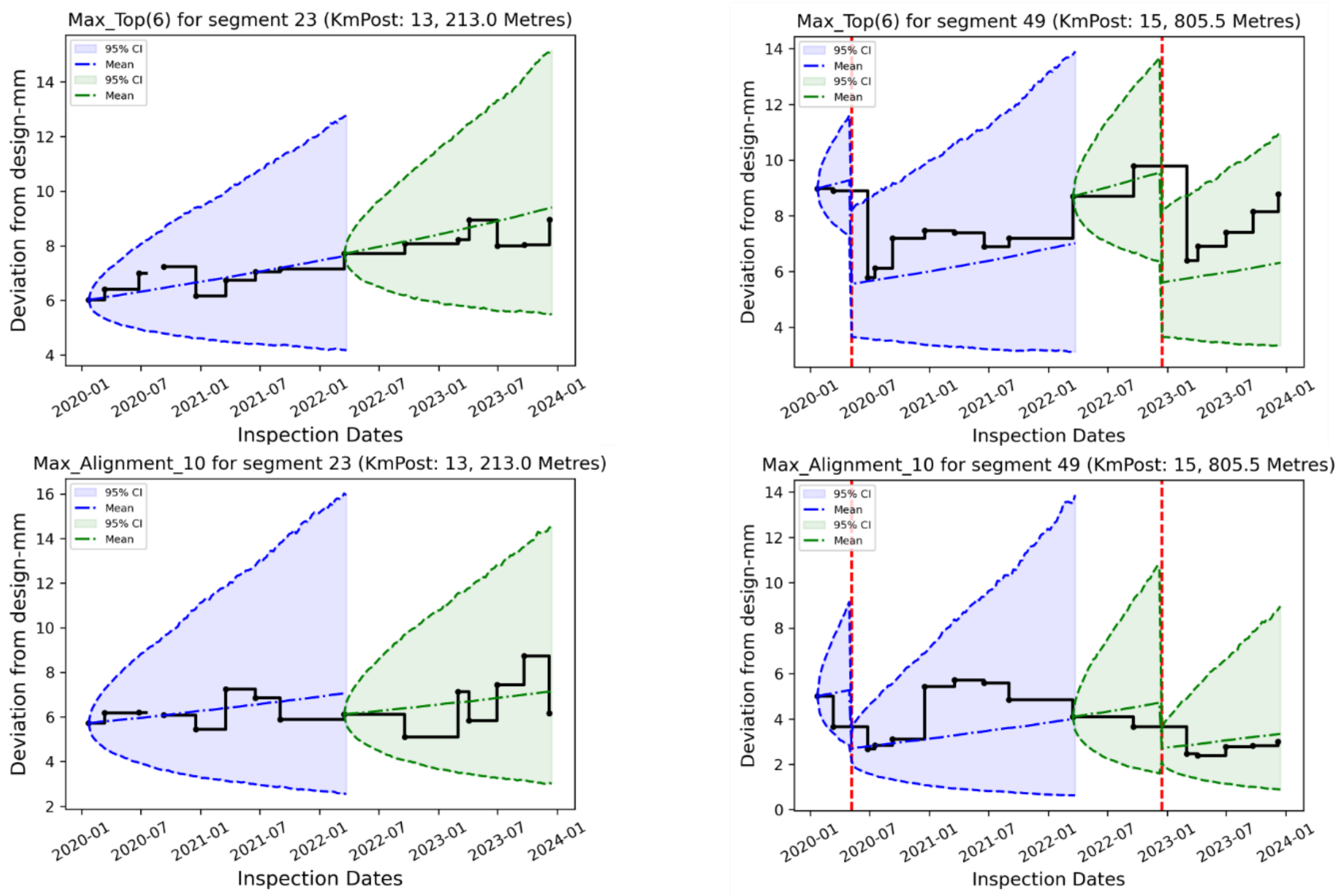

Fig. 21 Validation of log-space degradation model: without maintenance (left) and with maintenance (right)

**Fig. 22** displays the hitting time distributions for four maintenance thresholds—M3, M1, D7, and D1—using the log-space model of track geometry indicators. Compared to the model developed directly from the indicators (**Fig. 19** – 2 indicators), the peaks for M3 and M1 are lower and shift to the right. Conversely, the peak for D1 remains nearly the same in both models, but in the log-space model, it shifts to the left. In other words, the peaks of the hitting time distributions for the four maintenance thresholds are closer to each other in the log-space model. This is because the logarithmic transformation brings the thresholds closer together, leading to a shorter time for the degraded track geometry to reach D1.

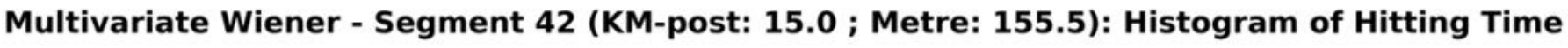

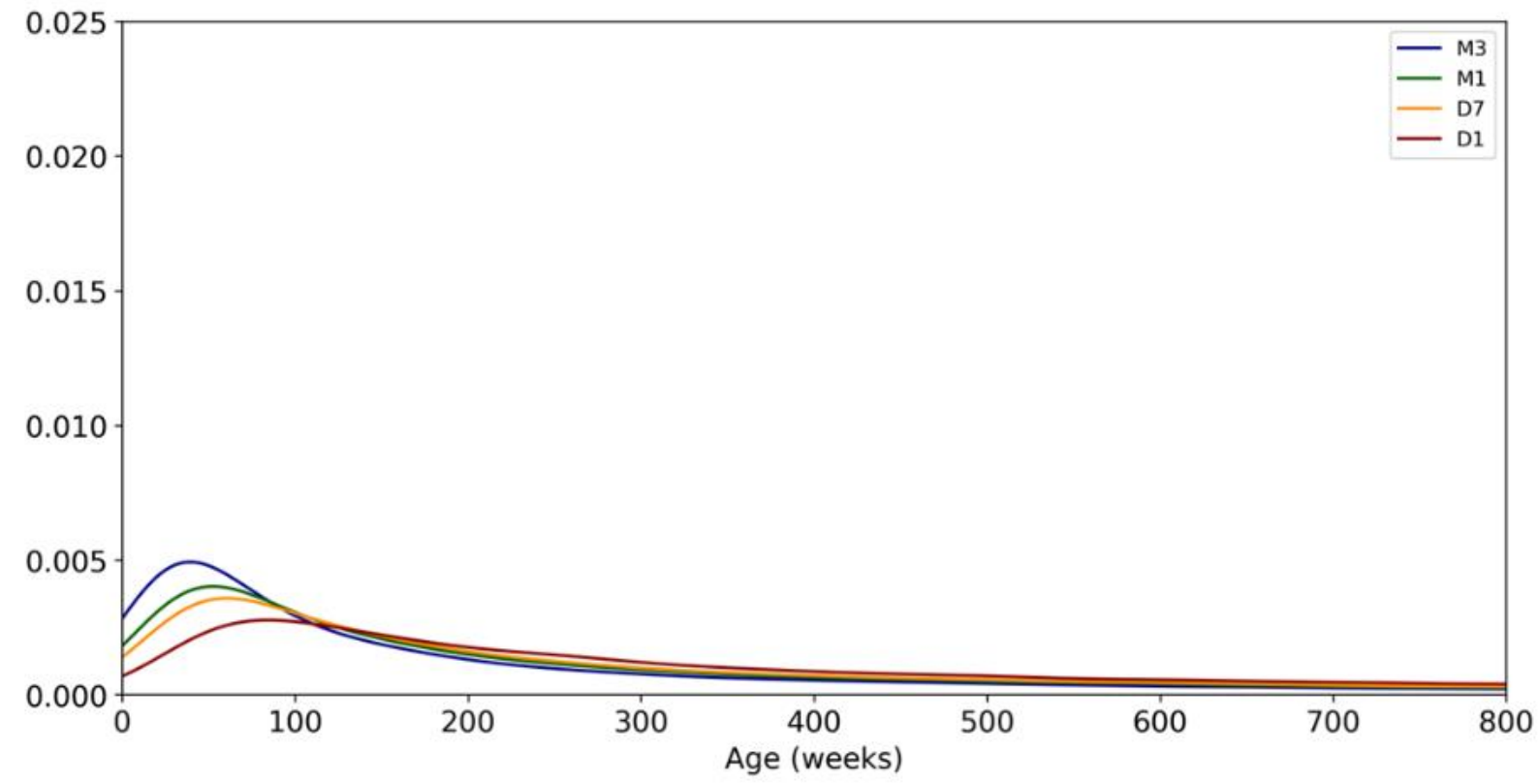

**Fig. 22** Modelling in log-space: Histogram of Hitting Time

# 5 Application of degradation model in inspection planning

As presented in Section 3, the track geometry inspection (i.e., TRV run) for the investigated track is planned at fixed intervals of 3 months. When any indicator reaches the maintenance thresholds M3, M1, D7, and D1, tamping is scheduled for that segment at intervals of 3 months, 1 month, 1 week, or 1 day, respectively. Thus, the threshold D1 can be considered the failure threshold of track geometry degradation. The developed degradation model above allows us to estimate the failure probability (i.e., reaching D1) under the current inspection strategy and develop a condition-based inspection policy for track geometry. This section will discuss the application of the degradation model in developing a condition-based inspection policy that can reduce the number of TRV runs while maintaining the same failure probability as the current inspection plan[2].

For this application, we interpret the thresholds as follows. If an M3 threshold is exceeded, it is assumed that maintenance (tamping) action will be taken before the next inspection. If M1 is exceeded, tamping will occur within one month, and if D7 is exceeded tamping will occur within the week. It is further assumed that the tamping takes place precisely at the middle of these intervals, i.e. for an M3 exceedance, tamping takes place at 1.5 months from the detected exceedance.

## 5.1 Failure probability of current periodic inspection and maintenance strategy

With the current 3-month periodic inspection strategy and the above maintenance policy, an upper bound on the probability of a D1 exceedance for a track (i.e. a collection of $N_q$ segments) may be computed as a conservative indication of the probability of a failure. This upper bound is computed as the probability that a particular segment reaches its D1 threshold before a planned maintenance or inspection occurs, given that its last monitored track degradation was close to (but smaller than) M3. Let $t$ be the most recent inspection time. The highest failure probability of segment $i$ is denoted as $p_{lim,i}$ and is defined as follows:

$$
\begin{aligned}
p_{lim,i} &= \mathbb{P}\left[\bigcup_{q=1}^{N_q} z_{M3,q} + \Delta Z_{q,i}(3) \geq z_{D1,q}\right] \\
&= \mathbb{P}\left[\bigcup_{q=1}^{N_q} \Delta Z_{q,i}(3) \geq z_{D1,q} - z_{M3,q}\right] \\
&= 1 - \mathbb{P}\left[\bigcap_{q=1}^{N_q} \Delta Z_{q,i}(3) \geq z_{D1,q} - z_{M3,q}\right] \\
&= 1 - F_{i,3}(\boldsymbol{z}_{D1} - \boldsymbol{z}_{M3})
\end{aligned} \tag{13}
$$

[2] Note that an alternative use could be to minimize the failure probability while maintaining the same utilization of the TRV.

Where $\Delta Z_{q,i}(d) = Z_{q,i}(t+d) - Z_{q,i}(t)$ is the change in indicator $q$ over time interval $d$ (in months), while $F_{i,d}(\Delta \boldsymbol{z})$ is the cumulative distribution function (CDF) of multivariate Normal distribution from the degradation model of segment $i$ with time interval $d$.

## 5.2 Condition-based inspection

Rail track geometry maintenance relies on condition-based maintenance, where tamping decisions are made based on measured track geometry indicators during inspections. The concept of condition-based inspection extends this approach to the deployment of the TRC. In the current periodic inspection strategy, if the measured degradation is below the first maintenance threshold (M3), the track segment will be inspected again in the next three months. In the condition-based inspection approach, the timing of the next inspection is also based on the measured degradation. If the measured indicator is high but still below M3, the track segment is scheduled for inspection in the next three months. However, if the degradation is low, a longer interval (e.g., four or five months) can be planned for the next inspection. **Fig. 23** shows an illustration of the condition-based inspection and maintenance for track geometry.

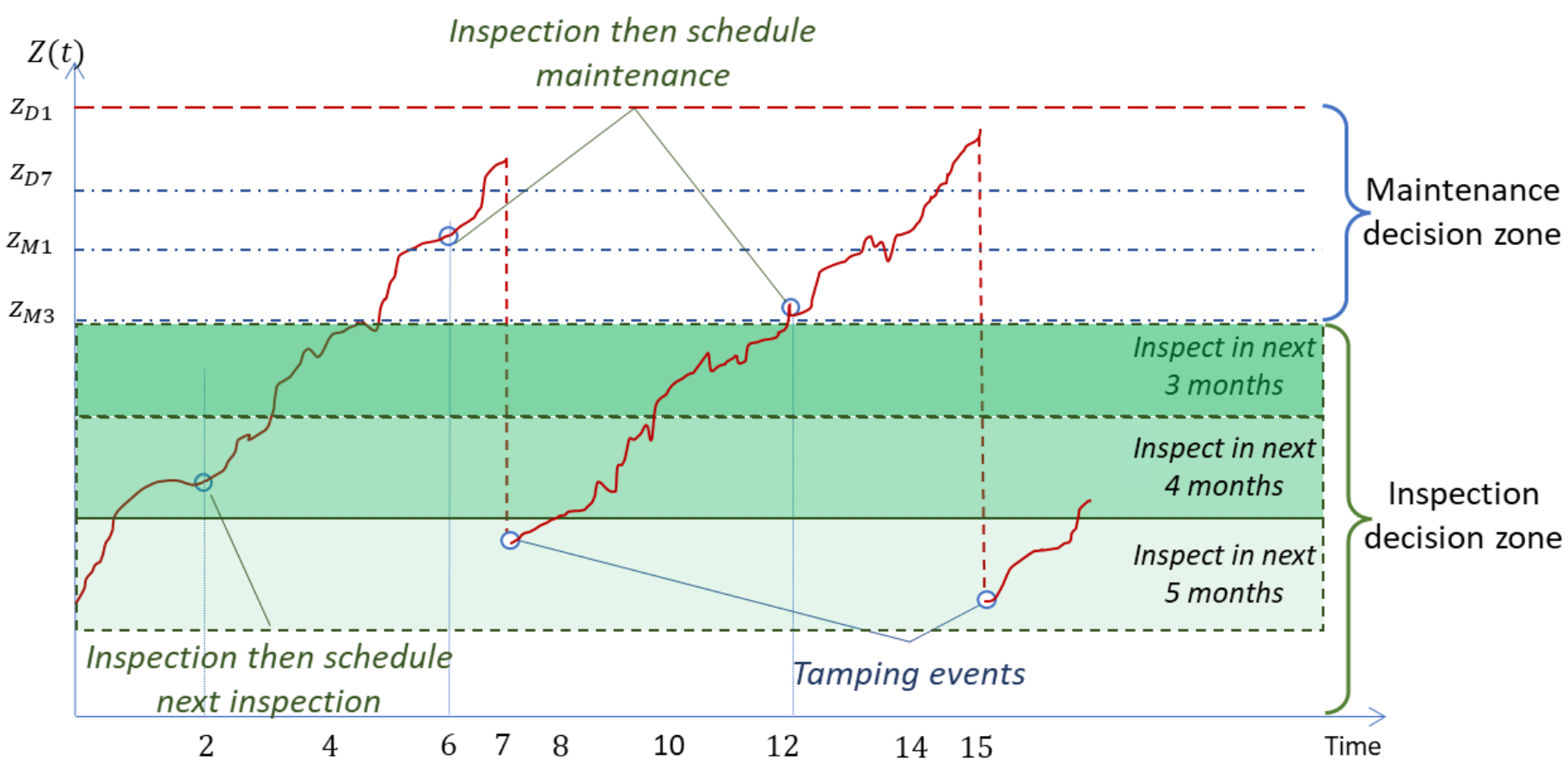

**Fig. 23** Illustration of condition-based inspection and maintenance for track geometry

Let $\boldsymbol{z}_d, d = 3,4,5, \ldots$, be the degradation threshold of all segments that the next inspection will be scheduled at $d$ months from current inspection time. Thus, $\boldsymbol{z}_{M3} \equiv \boldsymbol{z}_3 > \boldsymbol{z}_4 > \cdots$, and if the monitoring degradation $\boldsymbol{Z}(t)$ is in interval $[\boldsymbol{z}_{d+1}, \boldsymbol{z}_d)$, the next inspection will be set at $d$ months later. The degradation thresholds of inspection decision zones $\boldsymbol{z}_d$ should be determined through optimization modelling such as Markov Decision Process (MDP) [43] or some stochastic models. These models usually require an objective function derived from cost, benefit, or risk models of maintenance activities. However, for the current study on track geometry inspection and maintenance, we currently lack information on the cost or risk of track geometry degradation and maintenance expenses. Therefore, in this study, we estimate the inspection decision zone thresholds based on a user-defined acceptable failure probability $p_{lim,i}$, determined above.

For a specific segment $i$, let $\boldsymbol{z}_{d,i}$ be the threshold for this segment only. The upper bound for the probability that failure occurs during the inspection delay time may be computed by assuming that the most recent TRV run is very close to $\boldsymbol{z}_{d,i}$. Thus, the following inequality is satisfied:

$$P\left[\Delta \boldsymbol{Z}_i(d) \geq \boldsymbol{z}_{D1} - \boldsymbol{Z}_i(t) | \boldsymbol{z}_{d+1,i} \leq \boldsymbol{Z}_i(t) < \boldsymbol{z}_{d,i}\right] < P\left[\Delta \boldsymbol{Z}_i(d) \geq \boldsymbol{z}_{D1} - \boldsymbol{z}_{d,i}\right]$$

Let $p_{d,i}(\boldsymbol{z}_{d,i}) = P\left[\Delta \boldsymbol{Z}_i(d) \geq \boldsymbol{z}_{D1} - \boldsymbol{z}_{d,i}\right]$ be the representative probability of segment $i$ for the inspection decision zone $d$ months. This vector $\boldsymbol{z}_{d,i}$ is determined according to the searching process as following:

$$\boldsymbol{z}_{d,i} = \underset{\boldsymbol{z} \in \mathcal{D}_{\boldsymbol{z}}}{\operatorname{argmax}}\, \boldsymbol{z} \quad \text{s.t:}\; p_{d,i}(\boldsymbol{z}_{d,i}) = 1 - F_{i,d}(\boldsymbol{z}_{D1} - \boldsymbol{z}_{d,i}) \leq p_{lim,i} \;\; \forall d \tag{14}$$

Where $\mathcal{D}_{\boldsymbol{z}}$ is the searching domain of vector $\boldsymbol{z}_{d,i}$ which is from 0 to $z_{M3,q}$ for each indicator $q$. Then the threshold $\boldsymbol{z}_d$ for the track including all segments will be:

$$\boldsymbol{z}_d = \min_i \boldsymbol{z}_{d,i} \tag{15}$$

## 5.3 Results of condition-based inspection policy

For the illustration of the proposed condition-based inspection, the 2-indicator degradation model with Top and Alignment is used for the same track described in Section 3. The values of maintenance thresholds M3, M1, D7 and D1 are from the maintenance standards of the industry partner. The inspection and tamping of track are decided according to the monitored degradation levels of all segments. From the degradation models developed for each segment and Eq. (13) above, the $p_{lim,i}$ for each segment can be calculated. Then the degradation thresholds of inspection delay times are determined via the Eqs. (14) and (15). The results of inspection delay thresholds are shown in Table 3 below.

Table 3: Inspection delay thresholds

| Threshold | Top | Alignment |
|---|---|---|
| Next inspection in next 3 months | 14 | 21 |
| Next inspection in next 4 months | 11 | 18.5 |
| Next inspection in next 5 months | 9 | 16.5 |
| Next inspection in next 6 months | 6.5 | 14 |
| Next inspection in next 7 months | 4.5 | 12 |
| Next inspection in next 8 months | 3 | 10.5 |
| Next inspection in next 9 months | 1 | 8.5 |

With these thresholds, the timing for the next inspection is determined at the current inspection. For instance, if the Top and Alignment indicators during the current inspection of a specific segment show values of 8mm and 15mm, respectively, the next inspection for this segment will be scheduled in five months.

To evaluate the proposed condition-based inspection policy, the Monte Carlo simulation is used with 2,000 repetitions. **Fig. 24** shows the frequency of each inspection delay option compared to the benchmarked 3-month inspection plan. The 3-month delay remains the most frequent option. This is because the inspection delay decision applies to the entire track of 182 segments, where the condition of any segment can influence the inspection decision. Additionally, the values of track indicators after tamping are consistently high across all segments (**Fig. 4**). However, there is still several inspection delays that can make some savings for rail operators.

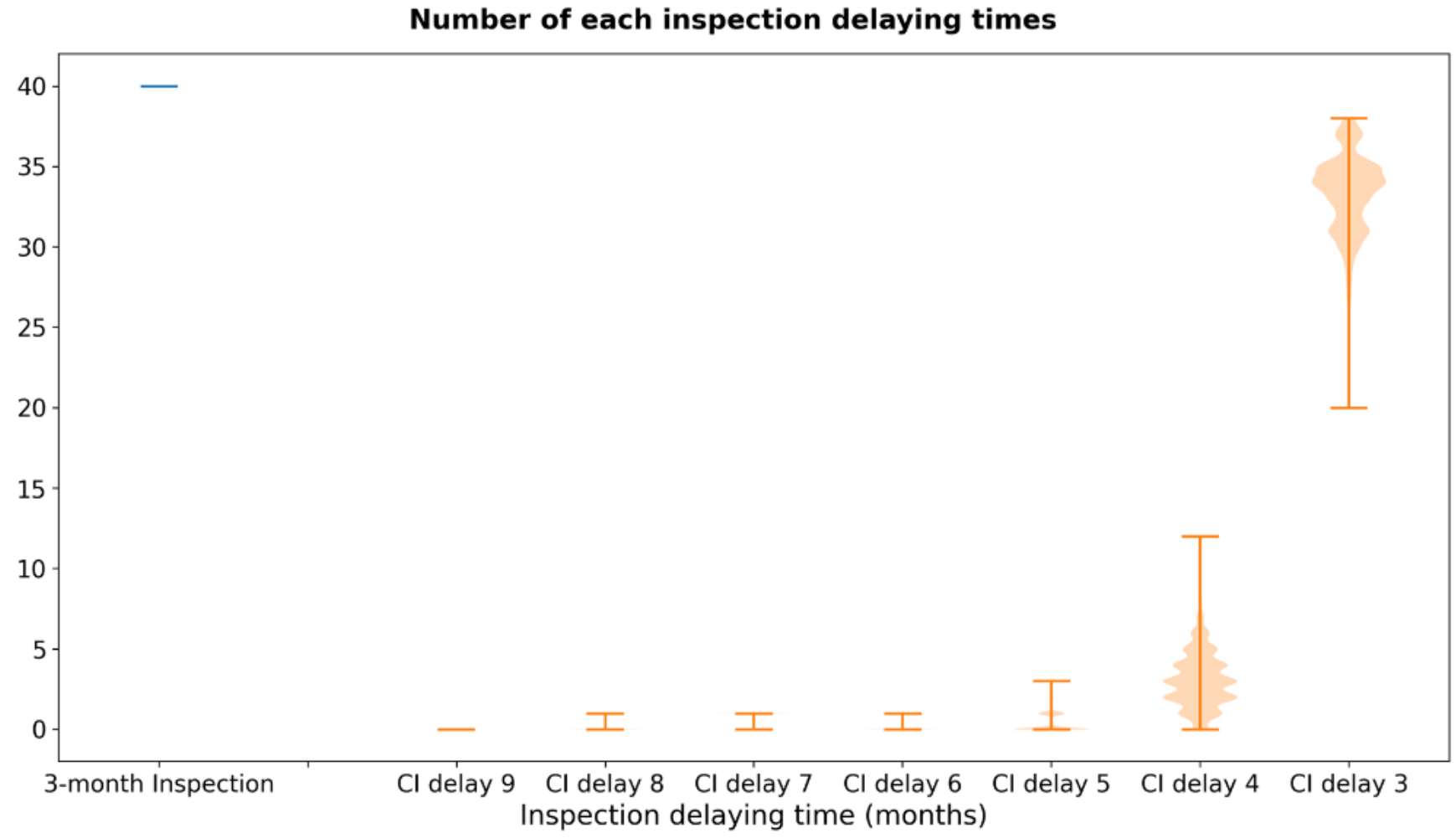

**Fig. 24** Number of inspections for each delay decision

**Fig. 25** below shows the outcomes from the simulation of two inspection policies: the 3-month periodic inspections and the condition-based inspections. As discussed above, the condition-based inspection policy can save a few TRV runs, approximately 2 to 3 times over 10 years, which is caused mostly by the high post-maintenance geometry indicators $\mathbf{z}_{i,k}^{+}$ for this track. In case tamping activities can yield better track geometry, a more significant reduction of the number of TRV runs is expected (see **Fig. 26**). The performance of the condition-based inspection policy is comparable to the benchmarked 3-month inspection plan, as indicated by the same number of maintenances and failures (i.e., hitting the D1 threshold).

The development of this condition-based inspection delay policy demonstrates the capacity and support of the developed multivariate degradation model. This degradation model is also useful for simulations that can test and validate any proposed inspection and maintenance policies from rail operators in the future.

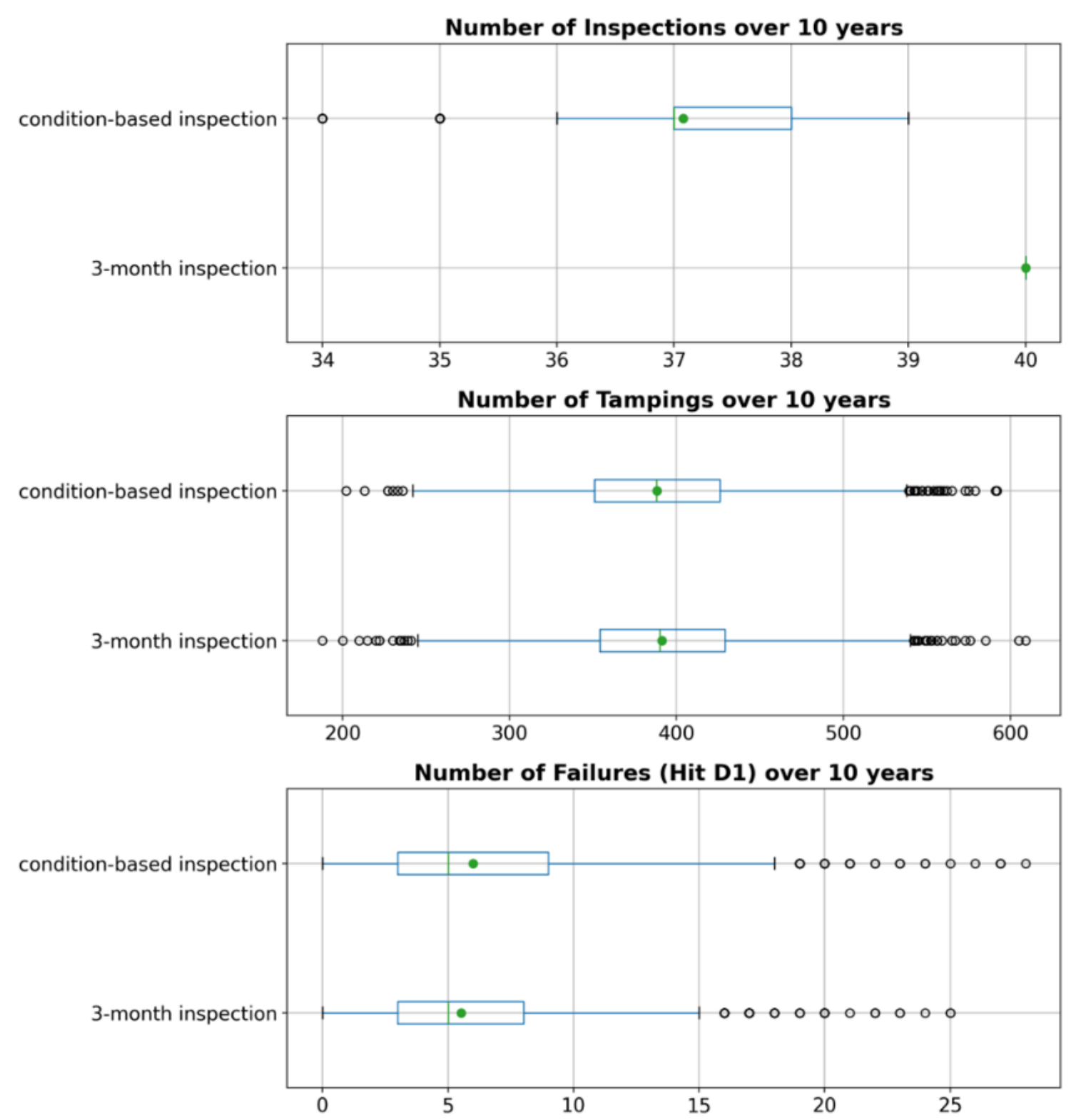

**Fig. 25** Numbers of inspection, maintenance, and failures of two inspection policies over 10 years

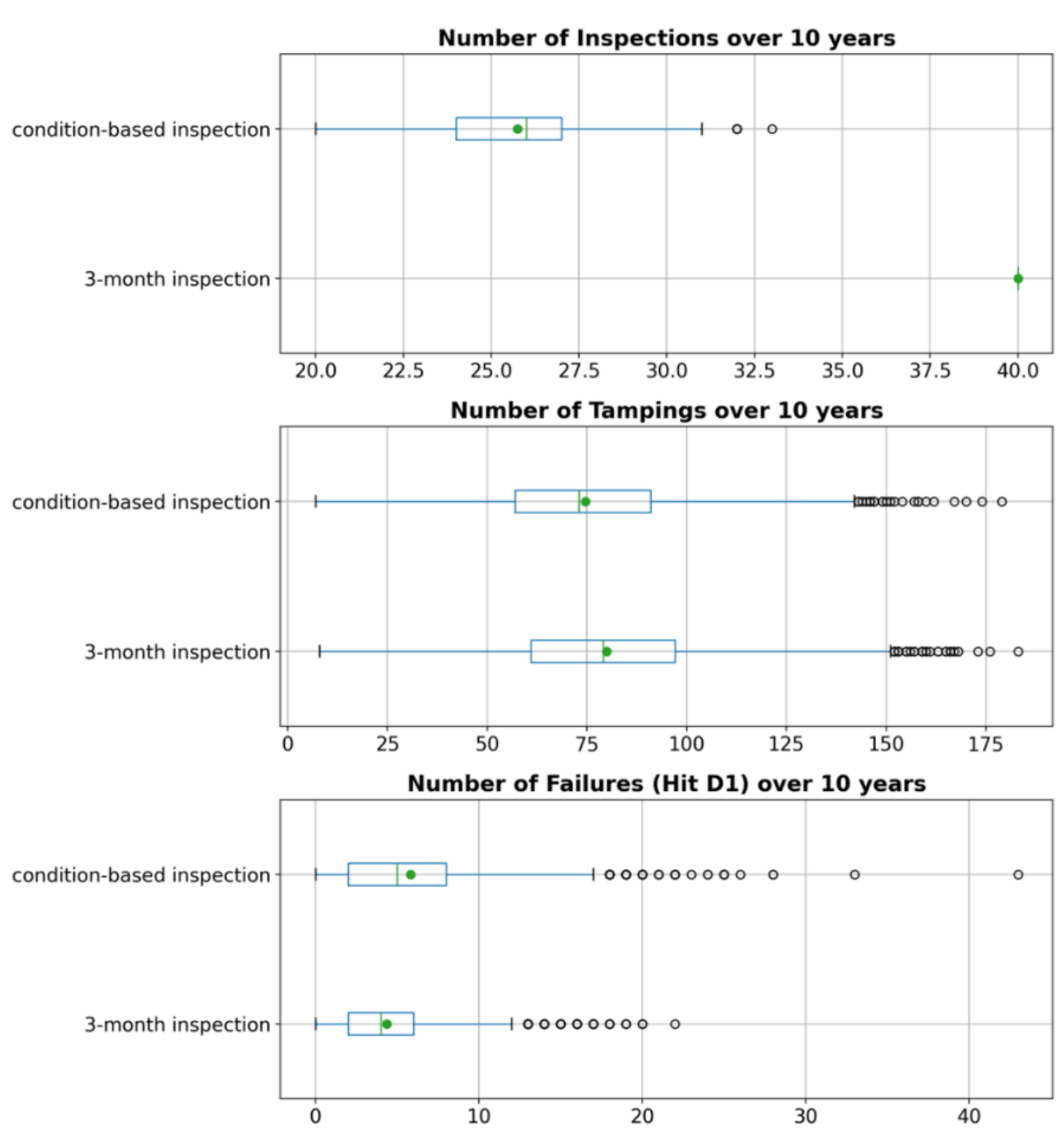

**Fig. 26** Numbers of inspection, maintenance, and failures of two inspection policies in case of lower track geometry indicators after tamping

# 6 Discussion and Conclusions

This study presents a novel methodology for modeling rail track geometry deterioration using a modified multivariate Wiener process. This approach captures correlations among multiple indicators and accounts for random geometry restoration due to imperfect maintenance. Bayesian inference enhances estimation accuracy for post-tamping conditions, while a hierarchical Bayesian approach further improves parameter estimation across track segments, reducing uncertainty (e.g., [0,0.01] vs. [0,0.118] for $\mu$ and [0.032,0.558] vs. [0.03,0.832] for $\sigma$ in the Wiener model).

The model was validated using real-world data from Queensland, Australia, demonstrating its advantages over single-indicator models, which tend to be overly conservative, leading to redundant inspections and excessive maintenance. By integrating multiple indicators, the proposed model improves track condition prediction, supporting more accurate and cost-effective inspection and tamping strategies.

The study finds that degradation is primarily driven by random variance rather than systematic drift, making condition-based inspection strategies particularly effective. A proposed inspection delay policy, tested against the standard 3-month cycle, shows potential for reducing TRV runs while maintaining inspection and maintenance performance.

While the model currently considers only longitudinal level and alignment, it can be extended to include other indicators such as gauge, cant, and twist. It does not yet incorporate external factors like speed, tonnage, soil conditions, or weather, which require additional data for analysis. However, the model is adaptable for global track networks with sufficient track geometry data.

Future research should explore the impact of environmental and operational factors on track degradation and develop optimization models for inspection scheduling, routing, and tamping planning to enhance railway maintenance efficiency.

## Acknowledgement

This study has been undertaken as part of the Australian Research Council (ARC) Linkage Project LP200100382.

## Statements and Declarations

The authors declare that they have no known competing financial interests or personal relationships that could have appeared to influence the work reported in this paper.